\documentclass[aps,prb,reprint,superscriptaddress]{revtex4-2}

\usepackage[english]{babel}                     % English
\usepackage[T1]{fontenc}                        % naprednejše kodiranje fonta
\usepackage[utf8]{inputenc}                     % pravilno razpoznavanje unicode znakov
\usepackage{amsmath,amssymb,amsfonts,amsthm}    % matematični paketi
\usepackage{commath}
\usepackage{url}                                % \url and \href for links
\usepackage{bm}                                 % bold math
\usepackage{graphicx}
\usepackage{hyperref}
\hypersetup{
    colorlinks=true,   % false: boxed links; true: colored links
    linkcolor=blue,    % color of internal links
    citecolor=blue, % color of links to bibliography
    filecolor=blue, % color of file links
    urlcolor=blue,     % color of external links
    runcolor=blue
}

\usepackage{times}
\usepackage[autostyle]{csquotes}
\usepackage{xcolor}
\usepackage{mathtools}
\usepackage{siunitx}

\usepackage{cleveref}

\crefname{appendix}{Appendix}{Appendices}
\crefname{equation}{Eq.}{Eqs.}
\crefname{figure}{Fig.}{Figs.}
\crefname{table}{Table}{Tables}
\crefname{section}{Sec.}{Secs.}
\Crefname{equation}{Equation}{Equations} % for the references in the beginning of sentences
\Crefname{figure}{Figure}{Figures}
\Crefname{section}{Section}{Sections}

\newcommand{\identity}{\ensuremath{\bm{1}}}         % identity operator
\newcommand{\iu}{\mathrm{i}}                            % imaginary unit
\newcommand{\e}{\mathrm{e}}                             % Euler's number
\newcommand{\mat}[1]{\ensuremath{\bm{\mathrm{#1}}}}     % matrices are bold
\renewcommand{\vec}[1]{\ensuremath{\bm{\mathrm{#1}}}}   % vectors are bold
\renewcommand{\Im}{\mathrm{Im}\,}                       % imaginary part

\DeclareMathOperator{\sign}{sign}
\DeclareMathOperator{\Tr}{Tr}
\DeclareMathOperator{\arcsinh}{arcsinh}

\makeatletter
\newsavebox{\@brx}
\newcommand{\llangle}[1][]{\savebox{\@brx}{\(\m@th{#1\langle}\)}%
  \mathopen{\copy\@brx\kern-0.5\wd\@brx\usebox{\@brx}}}
\newcommand{\rrangle}[1][]{\savebox{\@brx}{\(\m@th{#1\rangle}\)}%
  \mathclose{\copy\@brx\kern-0.5\wd\@brx\usebox{\@brx}}}
\makeatother

\begin{document}

\title{Strongly correlated Josephson junction: proximity effect in the single-layer Hubbard model}

\author{Don Rolih}
\email[]{don.rolih@ijs.si}
\affiliation{Jo\v{z}ef Stefan Institute, Jamova 39, 1000 Ljubljana, Slovenia}
\affiliation{Faculty of Mathematics and Physics, University of Ljubljana, Jadranska 19, 1000 Ljubljana, Slovenia}
%\thanks{}
% \altaffiliation{}
\author{Rok Žitko}
\email[]{rok.zitko@ijs.si}
\affiliation{Jo\v{z}ef Stefan Institute, Jamova 39, 1000 Ljubljana, Slovenia}
\affiliation{Faculty of Mathematics and Physics, University of Ljubljana, Jadranska 19, 1000 Ljubljana, Slovenia}

\date{\today}

\begin{abstract}
We study the proximity effect in the Hubbard model coupled to BCS superconductors describing a single-layer strongly correlated electron system in a phase-biased Josephson junction. We find two distinct gapped solutions, one Mott-like insulating (M-phase) and one proximitized superconducting phase (S-phase), separated by first-order transition with hysteresis. In the M-phase the large correlation charge gap strongly suppresses the critical current, while the S-phase behaves as a $0$-junction, with a proximitized gap that closes for $\phi=\pi$ to yield a correlated metal. Phase bias and junction transparency can thus serve as tuning knobs to switch between conducting and insulating regimes. Working within the dynamical mean field theory using the numerical renormalization group as the impurity solver, we associate M- and S-phase solutions with the doublet and singlet fixed points of the underlying superconducting Anderson impurity problem. We obtain detailed insight into the spectral structure on all energy scales. In the M-phase, the self-energy has sub-gap resonances symmetrically located around the Fermi level resulting from the splitting of the ``mid-gap pole'' found in Mott insulators; this structure accounts for phase insensitivity.
\end{abstract}

\maketitle

\section{Introduction}\label{sec:introduction}
Mott insulators, heavy-fermion metals, unconventional superconductors and other quantum materials show emergent phenomena that arise from strong electron-electron interactions \cite{keimer2017_physics_quantum_materials_NatPhys, tokura2017_emergent_functions_quantum_materials_NatPhys,balents2020moireflatbands}. Bringing together materials with competing ground states can create new electronic phases at the interfaces \cite{Boschker2017QuantumMatterHeterostructures}, motivating research on heterostructures combining superconductors with strongly correlated electron systems \cite{wang2024vdwjj,ma2025vdwscfm,dibbernardo2026vdwelectronics},
%where the key mechanism is the proximity effect: the penetration of electronic order from one material into another. 
%which for spinful states behave as $\pi$-junctions, where the current flows opposite to that in conventional junctions \cite{Glazman1989JETPLett,RozhkovArovas1999_PRL_2788,dam2006}. 
such as those based on van der Waals systems \cite{geim2013_vdW_heterostructures_Nature,wang2024vdwjj}, for example
%such as magic-angle twisted bilayer graphene \cite{cao2018_unconventional_superconductivity_magic_angle_graphen_Nature},
Nb$_3$Br$_8$, where one observes the Josephson diode effect \cite{wu2022_josephson_diode_vanderWaals_Nature,Dubbelman2025},
%and paramagnetic single-band Mott insulator state \cite{gao2023_mott_insulator_vanderWaals_Nb3Cl8_PhysRevX}, 
and two-dimensional heavy-fermion Kondo lattices grown on superconducting substrates \cite{kim2024_proximity_induced_superconductivity_2d_kondo_lattice_NanoLett}. 
Josephson junctions force the embedded correlated material to carry supercurrent whose magnitude and phase dependence encode information about the low-energy many-body states, similar to the situation in quantum dot devices with superconducting contacts \cite{RozhkovArovas1999_PRL_2788,dam2006,meden2019_anderson-jospehson_quantum_dot_theory_JourPhysCondMat}. The need to better understand such states motivates detailed studies of proximitized superconductivity in strongly-correlated phases. In this Letter, we address this question for a single-layer Mott insulator described by the Hubbard model. We will establish that a proximitized Mott insulator layer can be essentially Josephson-inactive: there exists a Mott solution branch with tiny phase stiffness (unlike in superconductor--band insulator--superconductor junctions with substantial Josephson energy), while the superconducting branch behaves as a conventional 0-junction whose proximitized gap closes at $\phi=\pi$, yielding a correlated metal.

\begin{figure}[hb]
  \includegraphics{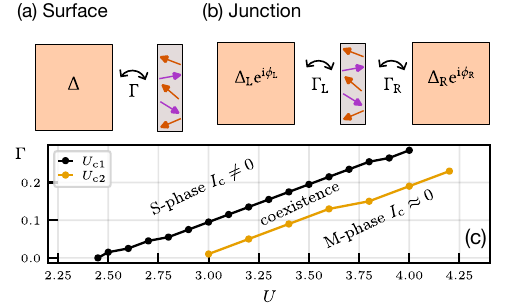}
  \caption{
    (a,b) Schematic representations of surface and Josephson junction problems. $\Delta$ and $\phi$ are the superconducting gap and phase, 
    $\Gamma$ are the tunneling hybridization strengths between correlated (gray) and superconducting regions (orange).
    (c) Phase diagram at half filling and zero temperature with superconducting (S) and Mott (M) phases. $U_{c1}$ and $U_{c2}$ are spinodal interaction strengths that bound the coexistence region. $I_c$ denotes the critical current.
  }
  \label{fig:fig1_device_phase_diagram_sketch}
\end{figure}

The Hubbard model is the paradigmatic model of strong correlation effects and describes the essential physics of the Mott metal-insulator transition (MIT) \cite{Imada1998MIT}.
Dynamical mean-field theory (DMFT) provides a non-perturbative framework for treating local correlations and captures Mott physics  \cite{georges1996_dmft_RevModPhys, qin2022_hubbard_model_computational_perspective_AnnuRevCondMattPhys}: 
a first-order phase transition occurs as a function of interaction strength with coexistence of solutions between interaction strengths $U_{c1}$ and $U_{c2}$ at low temperature\cite{GeorgesKrauth1992PRL,RozenbergZhangKotliar1992PRL,Georges1993,ZhangRozenbergKotliar1993PRL,SakaiKuramoto1994SSC,RozenbergKotliarZhang1994PRB_MH2}.
Extensions of the DMFT to inhomogeneous systems \cite{potthoff1999a_surface_metal_transition_Hubbard_PhysRevB} allow studies of layered heterostructures and interfaces \cite{Freericks2001, FREERICKS2002, miller2001, freericks2004_DMFT_inhomogeneous_multilayered_PhysRevB,okamoto2004_electronic_reconstruction_Mott_band_insulator, okamoto2004_inhomogeneity_strong_correlation_physics_DMFT_Mott_insulator_band_insulator_interface_PhysRevB,mazza2021_SC_heterostructures_DMFT_PhysRevB, petocchi2016_dmft_sc_layered_materials_PhysRevB}, for example how a metal placed in contact with a Mott insulator makes the metallic state penetrate into the insulator 
\cite{helmes2008_kondo_proximity_effect_metal_into_mott_insulator_PhysRevLett, zenia2009_appearance_fragile_fermi_liquids_finite_width_mott_insulators_idmft_PhysRevLett}.
DMFT reduces the bulk problem to a self-consistently defined quantum impurity problem, in our case the superconducting Anderson impurity model (SAIM). SAIM has been studied extensively as a model for quantum dot Josephson junctions \cite{choi2004_kondo_effect_josephson_current_quantum_dot_PhysRevB, karrasch2008_josephson_current_single_Anderson_impurity_BCS_leads_PhysRevB,  meden2019_anderson-jospehson_quantum_dot_theory_JourPhysCondMat}. In these systems, the ground state is determined by the competition between the Coulomb repulsion, Kondo screening, and the superconducting pairing. The resulting singlet-doublet quantum phase transition (QPT) in the impurity problem \cite{Bargerbos2022_PRXQuantum_030311} provides the conceptual underpinning for understanding the lattice problem studied in this work.

We consider two geometries sketched in \cref{fig:fig1_device_phase_diagram_sketch}(a,b): a surface problem with a single superconductor and a Josephson junction between two superconductors with phase bias $\phi=\phi_R-\phi_L$. For $\phi=0$, the two cases are equivalent. We will demonstrate the existence of two distinct phases: a Mott-like phase (M) with free local moments and a superconducting phase (S) with induced pairing correlations. The transition between them is first order, similar to both the singlet-doublet quantum phase transition in SAIM and to the MIT in Hubbard model. In fact, the two phase transitions are in direct correspondence in this system, which is one of our key results. 

\begin{figure*}[htbp!]
  \includegraphics{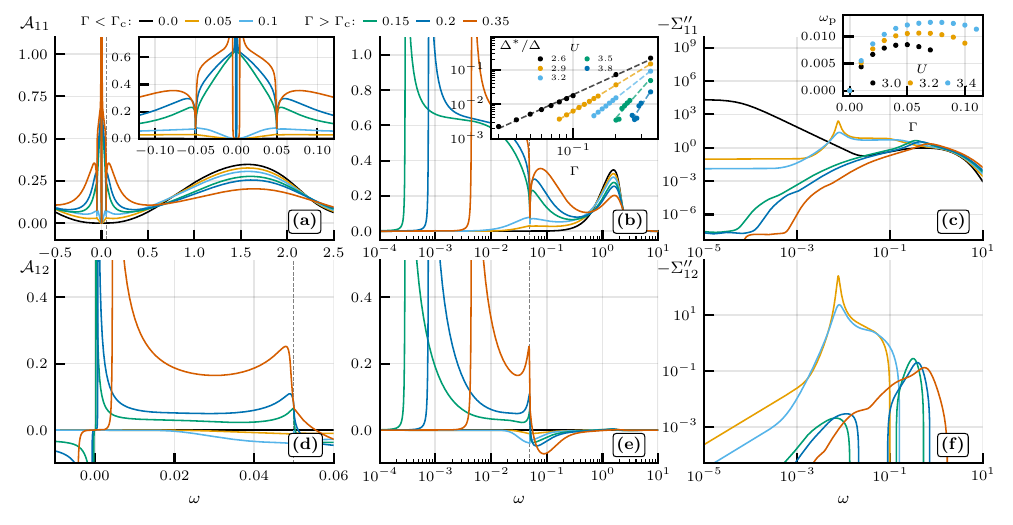}
  \caption{
    Normal (diagonal) spectral function $\mathcal{A}_{11}(\omega)$ on linear (a) and log scale (b), and anomalous (off-diagonal) spectral function 
    $\mathcal{A}_{12}(\omega)$ on linear (d) and log scale (e) at $U=3.2 > U_{c2}(\Gamma=0)$ for a range of $\Gamma$. The transition occurs for $\Gamma_c=0.13$. (c, f) Imaginary parts of the (c) normal and (f) anomalous components of the self-energy. Insets: (a) zoom into the low-frequency region, (b) induced gap $\Delta^{\ast}$ vs. $\Gamma$ for a range of $U > U_{c1}$ and (c) the position of the $\delta$ peak in the normal part of the self-energy vs. $\Gamma$.
  }
%  \label{fig:fig2_spectral_functions_Delta_star_tau_suscep_U=3.2_vs_Gamma}
 \label{fig2}
\end{figure*}

The paper is organized as follows. In \cref{sec:model_method} we introduce the Hubbard model coupled to superconducting regions and the DMFT self-consistency procedure. In \cref{subsec:surface} we study the surface problem and in \cref{subsec:junction} we present results for a strongly correlated Josephson junction. \Cref{sec:discussion} contains the discussion. Additional numerical results and technical details of the calculation can be found in the Appendices.

\section{Model and method}\label{sec:model_method}

The system is composed of a strongly-correlated region weakly coupled to two SC regions.
The strongly-correlated region is modeled by the Hubbard Hamiltonian
\begin{equation}
  \begin{split}
  H_{\mathrm{scr}} &= -t \sum_{\langle ij \rangle, \sigma} (d_{i\sigma}^\dagger d_{j\sigma} + \mathrm{h.c.}) - \mu \sum_{i,\sigma} d_{i\sigma}^\dagger d_{i\sigma}\\ 
  &+ U \sum_{i} n_{i\uparrow}n_{i\downarrow},
  \end{split}
\end{equation}
where $d_{i\sigma}^\dagger$ creates an electron of spin $\sigma$ on site $i$ of the Bethe lattice, $t$ is the nearest-neighbor hopping parameter, $\mu$ is the chemical potential, and $U>0$ is the on-site Coulomb repulsion. The Bethe lattice non-interacting density of states (DOS) is given by
\begin{equation}
  \rho(\epsilon) = \frac{\sqrt{4t^2 - \epsilon^2}}{2\pi t^2}, \quad |\epsilon| < 2t.
\end{equation}

The coupling to SC regions $\lambda=\mathrm{L}, \mathrm{R}$ for each Bethe lattice site $i$ of the Hubbard model is described by the Hamiltonian
\begin{equation}
  H_{\mathrm{sc}, \lambda} = V_{\lambda} \sum_{i, \vec{q},\sigma} [(d_{i\sigma}^\dagger c_{i\lambda, \vec{q} \sigma} + \text{h.c.}) + \sum_i H^{\mathrm{BCS}}_{i,\lambda}[c_{i\lambda, \vec{q}\sigma}],
\end{equation}
where the superconducting regions are described by the BCS Hamiltonian
\begin{equation}
  \begin{split}
  H_{i,\lambda}^{\mathrm{BCS}} &= \sum_{\vec{q},\sigma} \varepsilon_{\vec{q}} c_{i\lambda,\vec{q}\sigma}^\dagger c_{i\lambda,\vec{q}\sigma} \\
  &+ \sum_{\vec{q}} \left( \Delta_{\lambda} c_{i\lambda,\vec{q}\uparrow}^\dagger c_{i\lambda,-\vec{q}\downarrow}^\dagger + \mathrm{h.c.} \right).
  \end{split}
\end{equation}
Here $\varepsilon_{\vec{q}}$ is the dispersion relation in the superconductor, $\Delta_{\lambda} = |\Delta_{\lambda}|\e^{\iu\phi_{\lambda}}$ is the SC order parameter with amplitude $|\Delta_{\lambda}|$ and phase $\phi_{\lambda}$.
We consider the symmetric case $V = V_{\mathrm{L}} = V_{\mathrm{R}}$ and $\Delta = \Delta_{\mathrm{L}} = \Delta_{\mathrm{R}}$. 

Each $H_{i,\lambda}^{\mathrm{BCS}}$ acts as a reservoir for each site $i$ of the Hubbard model. 
An analogous approach has recently been used for the Hubbard model coupled to normal electron reservoir~\cite{bag2024_coupling_strongly_correlated_electron_system_elecetronic_reservoir_PhysRevB}, and it can be shown that 
this formulation is fully equivalent to the inhomogeneous DMFT approach for a single active layer coupled to semi-infinite leads~\cite{zenia2009_appearance_fragile_fermi_liquids_finite_width_mott_insulators_idmft_PhysRevLett,helmes2008_kondo_proximity_effect_metal_into_mott_insulator_PhysRevLett} (see Appendix~\ref{sec:iDMFT} for details).

We tackle this problem using the DMFT approach with extensions for superconducting order \cite{georges1996_dmft_RevModPhys,toschi2005_bcs_bec_dmft_attractive_Hubbard_model_PhysRevB, bauer2009_DMFT_NRG_hubbard_sc_PhysRevB}. The effective hybridization function of the impurity model in Nambu matrix space is
\begin{equation}
  \label{eq:Delta_eff}
  \mat{\Delta}_{\mathrm{eff}} = t^2 \mat{\tau}_3 \mat{G}_{\mathrm{loc}} \mat{\tau}_3 + \mat{\Sigma}_{\mathrm{BCS}},
\end{equation}
where $\mat{\Sigma}_{\mathrm{BCS}} = \sum_{\lambda} V^2 \mat{\tau}_3 \mat{g}_{\lambda}(z) \mat{\tau}_3$ is the self-energy due to the proximity effect, with $\tau_3$ the Pauli $Z$ matrix and $\mat{g}_\lambda$ the BCS Green's functions (see Appendix~\ref{sec:green_functions} for details of Green's functions definition, and Appendices~\ref{sec:self-consistency}, \ref{sec:hilbert} for details of the implementation of the DMFT self-consistency). 
We neglect the back-action of the correlated region on the baths (order-parameter suppression), which is appropriate for macroscopically large baths in contact with a single layer as a first approximation.
We focus on the paramagnetic solution at half-filling. Magnetic order would introduce an additional scale, but would not influence the phenomena central to this work---dynamical correlation effects, critical current suppression in the M-phase and metallization in the S-phase---since these are due to local self-energy physics that will survive upon magnetic ordering. All results will be given in energy units of half-bandwidth $D=2t$. The values of critical $U_{c1}$ and $U_{c2}$ for the Bethe lattice at $T = 0$ were estimated as $2.39$ and $2.92$ ~\cite{bulla1999_zero_temperature_metal_insulator_transition_nrg_PhysRevLett,zitko2013_extremely_correlated_fermi_liquid_DMFT_doping_driven_mott_PhysRevB,eisenlohr2019_mott_criticality_one_band_hubbard_DMFT_NRG_PhysRevB}. In all calculations, $\Delta=0.05$. Hybridization with the BCS contacts is quantified by $\Gamma=2\pi \rho_\mathrm{SC}(0)V^2$, where $\rho_\mathrm{SC}$ is the density of states in the SC baths.

To solve the effective impurity problem, we use the full-density matrix numerical renormalization group method~\cite{wilson1975_RG_Kondo_RevModPhys, bulla2008_nrg_RevModPhys, anders2005_real_time_dynamics_qis_nrg_many_body_basis_PhysRevLett, weichselbaum2007_sum_rule_conserving_spectral_functions_NRG_PhysRevLett}, specifically the NRG Ljubljana implementation~\cite{zitko2009_discretization_artefacts_PhysRevB, zitko2021_nrg_ljubljana},
employing the discretization scheme of Ref.~\cite{liu2016_channel_mixing_PhysRevB} and the improved self-energy estimators of Ref.~\cite{kugler2022_self_energy_trick_PhysRevB}.
We compute the results at $T=\num{e-8} \ll \Delta$. Residual in-gap spectral weight is a controlled NRG artifact due to broadening and clipping; see Appendix~\ref{sec:dep_nrg_params}, Figs.~\ref{fig:fig11_temperature_clip_dependence}, \ref{fig:fig12_nrg_parameters_dependence} and \ref{fig:fig13_Sigma_delta_peak_broadening}, for benchmarks.

\section{Results}
\subsection{Surface problem}\label{subsec:surface}
We start with the surface problem, \cref{fig:fig1_device_phase_diagram_sketch}(a), with $U=3.2>U_{c2}$ so that for $\Gamma=0$ we recover the standard Hubbard model in the Mott phase. In \cref{fig2}(a,b,d,e), we show the normal (diagonal) and anomalous (off-diagonal) local spectral function for a range of $\Gamma$ \footnote{
Spectral functions are defined as
$\mathcal{A}_{11}(\omega)=(-1/\pi)\mathrm{Im}\langle\langle d_\sigma;d_\sigma^\dag \rangle\rangle$
and
$\mathcal{A}_{12}(\omega)=(-1/\pi)\mathrm{Im}\langle\langle d_\downarrow;d_\uparrow \rangle\rangle$,
where $\llangle A; B \rrangle$ denotes the retarded correlation function for operators $A$ and $B$.
}. 
At $\Gamma=0$, we find the usual Mott gap between the Hubbard bands centered at $\pm U/2$. For small non-zero $\Gamma$ we observe spectral weight transfer from the Hubbard bands to energies around $\Delta$. The spectral function drops to zero continuously at the inner gap edges near $\Delta$, with no coherence peaks, similar to the behaviour at the Mott gap edges for $\Gamma=0$. The effective spectral gap is thus reduced from the Mott gap $\approx U-2D$ to a scale set by $\Delta$, which is best observed on the logarithmic scale in panel (b). The anomalous spectral function is negative for $\omega>0$, with spectral weight concentrated in the region around $\Delta$. All spectral functions exhibit cusps at $|\omega|=\Delta$, which originate from the square root singularity in $\mat{\Sigma}_{\mathrm{BCS}}$.

When $\Gamma$ is increased further, a discontinuous change occurs at some critical value $\Gamma_{\mathrm{c}}$ (for $U=3.2$ at $\num{0.13}$). A proximity-induced SC gap $\Delta^{\ast}$ with divergences at its edges (coherence peaks) appears in both normal and anomalous spectra. The spectral functions thus resemble those of a BCS superconductor with a gap $\Delta^{\ast}$ while retaining the high-energy Hubbard bands located around $\pm U/2$. As before, all spectra have cusps at $\Delta$. The induced gap $\Delta^*$ starts at a finite value at $\Gamma=\Gamma_\mathrm{c}$, then grows as a power-law $\Delta^* \propto \Gamma^\eta$ with $U$-dependent exponent $\eta$ \footnote{The values of $\eta$ are $1.89$, $2.52$, $3.25$, $4.72$, $6.34$ corresponding to values of $U$ shown in the inset \cref{fig2}(b) from smallest to largest.} as shown in the inset of \cref{fig2}(b). The bath coupling thus destroys the Mott state at $\Gamma=\Gamma_{\mathrm{c}}$ and immediately induces SC coherence peaks. In other words, the emergence of a metallic solution from the collapsing Mott state is preempted by the opening of a SC gap.
At large $\Gamma$, $\Delta^{\ast}$ tends toward $\Delta$. When passing through $\Gamma=\Gamma_c$, the anomalous spectral function has a discontinuous sign change in the range $|\omega|<\Delta$, see panels (d,e). The solution found for $\Gamma < \Gamma_c$ will be denoted as the M-phase, because it strongly resembles the Mott solution to which it smoothly connects in the $\Gamma\to0$ limit. The solution for $\Gamma>\Gamma_c$ will be denoted as the S-phase, due to superconducting coherence peaks in its spectra.

In \cref{fig2}(c, f), we relate local spectral functions to the corresponding self-energies. The key observation is that in the M-phase, the imaginary part of self-energy $\Im\Sigma$ shows pronounced peaks inside the $\Gamma=0$ Mott gap, with equal weight and position in both diagonal and out-of-diagonal components. These peaks arise from the characteristic Mott insulator $\delta$ peak at $\omega=0$ (the ``midgap pole'') that generates the Mott gap \cite{RozenbergKotliarZhang1994PRB_MH2,bulla1999_zero_temperature_metal_insulator_transition_nrg_PhysRevLett,Logan2015,Sen2020}. This peak splits for finite $\Gamma$ into two resonances that are symmetrically located around the Fermi level at energy $\omega_p$, see the inset to \cref{fig2}(c). With increasing $\Gamma$ they broaden and $\omega_p$ reaches a maximum value. The mechanism that drives the collapse of the Mott phase is that $\mathrm{Re}\Sigma$ at the Fermi level is no longer large enough to fully push out the spectral density. The transition is discontinuous and the subgap resonances in $\Im\Sigma$ disappear.
In the S phase, $\Im\Sigma$ is then zero for low $\omega$ within the induced gap (the finite noisy values in the $\omega\to0$ range are an artifact of the method).

\begin{figure}[t]
  \includegraphics{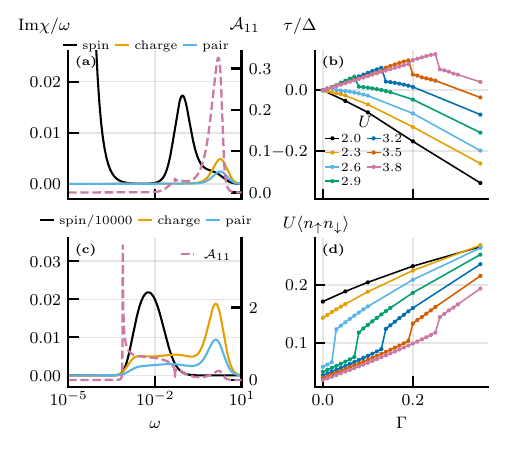}
  \caption{
    (a) Local susceptibilities for $\Gamma=0.05$ and (c) $\Gamma=0.2$ with $U=3.2$. We also show the diagonal component of the spectral function $\mathcal{A}_{11}$ for reference. (b) Local pair expectation value $\tau = U \langle d^{\dagger}_{\uparrow} d^{\dagger}_{\downarrow}\rangle$ and (d) double occupancy $U \langle n_{\uparrow} n_{\downarrow} \rangle$ vs. $\Gamma$.
  }
  \label{fig3}
\end{figure}

 Further differences are found in two-electron response. In \cref{fig3}(a, c), we show local charge, spin and pairing susceptibility 
\footnote{Local charge, spin and pairing susceptibility are defined as $\chi_{\mathrm{charge}}(\omega) = \llangle n_d; n_d \rrangle$, $\chi_{\mathrm{spin}}(\omega) = \sum_{\alpha}\llangle S^z; S^z \rrangle$, $\chi_{\mathrm{pair}}(\omega) = \llangle d_{\downarrow} d_{\uparrow}; d_{\uparrow}^\dagger d_{\downarrow}^\dagger \rrangle$, where $n_d = n_{d\uparrow} + n_{d\downarrow}$ and $S^z = 1/2(n_{d\uparrow} - n_{d\downarrow})$.}. The M-phase is characterized by an $\omega=0$ peak in the spin susceptibility. This is the signature of a well-defined local moment and is qualitatively similar to the DMFT spin susceptibility of the standard Hubbard model in the paramagnetic Mott phase. Charge and pairing susceptibilities are suppressed in the Mott gap. In the S-phase, however, the susceptibilities are non-zero only above the induced gap $\Delta^{\ast}$ and there is no $\omega=0$ peak in $\chi_\mathrm{spin}$. Additional results for the spectral function and susceptibility are shown in Appendix~\ref{sec:additional_surface}.

\begin{figure}[h!]
  \includegraphics{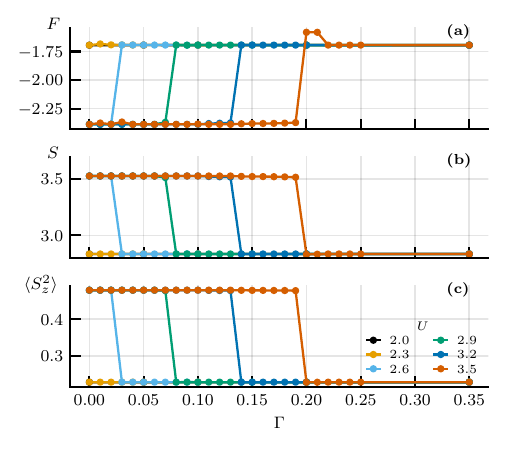}
  \caption{
    Thermodynamic observables (free energy $F$, entropy $S$, spin $S_z^2$) of the effective impurity model at DMFT self-consistency as a function of $\Gamma$ for $U=3.2$ starting from an insulating initial guess. Discontinuity in entropy is $\ln 2$, jump in spin squared is $1/4$.
  }
  \label{fig:thermodynamic_observables_insulator_U=3.2_sweep_Gamma}
\end{figure}

The expectation value of the local pairing operator $\tau=U\langle d^\dag_\uparrow d^\dag_\downarrow\rangle$ (i.e., defined as the pairing Hartree shift) is non-zero for finite $\Gamma$, including in the M-phase, see \cref{fig3}(b). Surprisingly, the transition does not correspond to a sign change of the pairing expectation value, but to a sign change of its first derivative (slope) as a function of $\Gamma$. This is unusual, because in the singlet-doublet QPT in SAIM this expectation value changes sign.  Discontinuities are also found in other local observables, such as double occupancy \cref{fig3}(d), and in thermodynamic quantities, which are consistent with a sudden disappearance of local moments. The latter are shown in \cref{fig:thermodynamic_observables_insulator_U=3.2_sweep_Gamma}: the free energy $F$, the entropy $S$, and the expectation value of the local moment squared of the effective impurity model at self-consistency for $U=3.2$. The results are consistent with the disappearance of the local moment for $\Gamma > \Gamma_{\mathrm{c}}$.

This behavior is suggestive of a first-order QPT between states that are connected with the $\Gamma=0$ phases of the Hubbard model. The $U_{\mathrm{c}1}$ Mott transition point can thus be continued from $\Gamma=0$ to finite $\Gamma$ in a well-defined way. Starting from an S-type initial state, we can define $U_{\mathrm{c}2}$ in a similar way. Between $U_{c1}$ and $U_{c2}$ we find a coexistence region, which extends from the $\Gamma=0$ points into the $(U,\Gamma)$ plane as two parallel straight lines in the range of $\Gamma$ shown, see Fig.~\ref{fig:fig1_device_phase_diagram_sketch}(c). The coexistence region needs to persist to arbitrarily large values of $\Gamma$, because the M and S phases are distinct at $T=0$. (With increasing $T$, the separation between the lines shrinks and disappears at the critical end point of the $\Gamma=0$ model.)
The linear form of the phase boundaries can be understood by considering the atomic limit of the problem, which is the same as in SAIM, i.e., a single interacting level with on-site interaction $\tilde{U}$ and pairing interaction $\tilde{\Delta} \propto \Gamma$. The lowest energy is $-\tilde{U}/2$ in the doublet subspace and $-\tilde{\Delta}$ in the singlet subspace, thus the transition line is given by $\tilde{U}/2 = \tilde{\Delta}$. We can thus understand M to S transition as a competition between the effective on-site interaction $U$ and the induced superconducting pairing.

\begin{figure*}[t!]
  \includegraphics{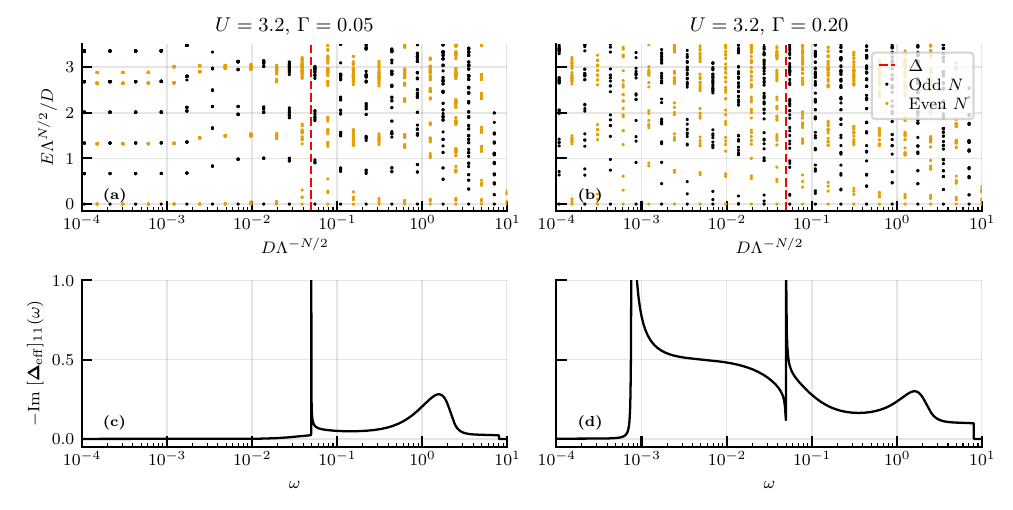}
  \caption{
    NRG renormalization group flow diagrams of the effective impurity model at DMFT self-consistency for $U=3.2$ and $\phi=0$ and two different values of the coupling to the superconducting region: $\Gamma = 0.05 < \Gamma_c$ in the M-phase (left column) and $\Gamma = 0.20 > \Gamma_c$ in the S-phase (right column).
    Bottom panels (c, d) show the effective hybridization function at DMFT self-consistency for both cases.
    The value of the superconducting gap is $\Delta = 0.05$. Here we use NRG discretization parameter $\Lambda = 2$ (instead of $\Lambda=4$) for better resolution.
  }
  \label{fig:rg_flow_U=3.2_Gamma=0.05_Gamma=0.20_Lambda=2}
\end{figure*}

In \cref{fig:rg_flow_U=3.2_Gamma=0.05_Gamma=0.20_Lambda=2}, we show the NRG renormalization group flow diagrams in the M and S phases. A different many-body spectrum in the two cases confirms that the phases are indeed distinct and associated with the two different fixed points of the superconducting Anderson impurity model (the well-known "doublet" and "singlet" ground states). Bottom panels show the hybridization function to emphasize that these are highly structured and vastly more complex that those that appear in the conventional SAIM modes for quantum dot Josephson junctions, which are taken to be featureless.

\subsection{Josephson junction problem}\label{subsec:junction}
\begin{figure}[t]
\includegraphics{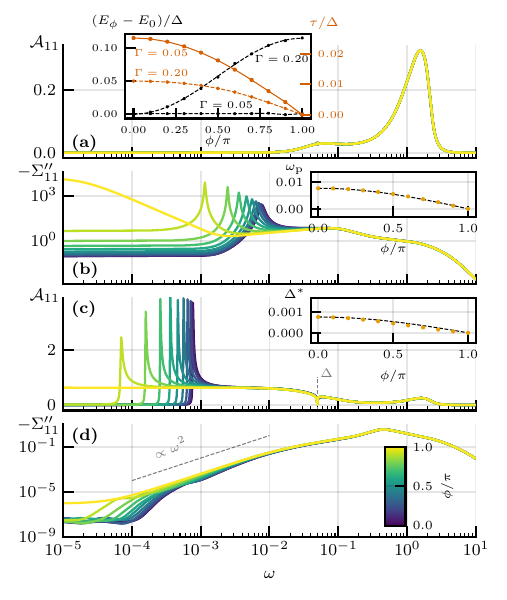}
  \caption{
    Phase dependence of the spectral function $\mathcal{A}_{11}(\omega)$ and the self-energy at fixed $U=3.2$ for (a, b) $\Gamma=0.05$ (M-phase) and (c, d) $\Gamma=0.2$ (S-phase). Note the Fermi-liquid $\omega^2$ scaling of the self-energy for small $\omega$ at $\phi=\pi$. Insets: (a) $\phi$-dependence of total energy $E(\phi)$ with cosine fit (dashed line) and the local pairing expectation value $\tau$; (b) position of the $\delta$ peak in the self-energy as a function of $\phi$; the dashed line is $\omega_\mathrm{p}^{\phi=0}\cos(\phi/2)$; (c) Induced coherence peak as a function $\phi/\pi$;  the dashed line is $\Delta^{\ast,\phi=0}\cos(\phi/2)$.
  }
  \label{fig:fig4}
\end{figure}

We now turn to the case of a Josephson junction with a finite phase bias $\phi$, represented in \cref{fig:fig1_device_phase_diagram_sketch}(b).
The superconducting regions couple to the Hubbard layer via $\mat{\Sigma}_{\mathrm{BCS}}$, which contains an off-diagonal component proportional to $\cos(\phi/2)$.
In \cref{fig:fig4}, we show the local spectral function and the self-energy at fixed $U=3.2$ in the M-phase (a, b) and in the S-phase (c, d) for the full phase bias range, $\phi\in[0, \pi]$.

In the M-phase, the spectral function shows no discernible dependence on $\phi$. The position $\omega_{\mathrm{p}}$ of the $\delta$ peak in the self-energy is proportional to the effective pairing strength and has a $\cos(\phi/2)$ dependence [inset \cref{fig:fig4}(b)]. Its weight, however, is independent of phase. At $\phi = \pi$, the coupling of the anomalous component to the superconductor vanishes and the pole returns to zero.
Although the self-consistency condition still induces a non-zero anomalous expectation value which depends on $\phi$, see the inset in \cref{fig:fig4}(a), the low-energy region of the spectrum $\omega \lesssim \Delta$ has no available states on which the pairing field could act. The charge gap arises from strong correlations rather than band structure, i.e., the gap is sustained by strong self-energy resonances rather than a single-particle band gap, and the resulting local moment remains completely decoupled from the superconducting regions. This is a local effect that is well captured by DMFT.
There are no low-energy quasiparticle or subgap states to support coherent pair transfer and the Josephson coupling is suppressed to tiny values (see Fig.~\ref{fig:energies_vs_phi_U_3.2_gap_0.05} for an upper boundary estimation: at least an order of magnitude suppression).
This is in contrast to conventional SIS tunnel junctions, where Cooper pair tunneling through a band insulator still produces a $\phi$-dependent ground state energy. It is also in contrast to the behavior of the associated magnetic (doublet) solution in the quantum dot problem which corresponds to the reversal of Josephson current \cite{Glazman1989JETPLett,RozhkovArovas1999_PRL_2788,dam2006}, rather than its suppression.
This almost complete insensitivity to phase bias in the M-phase is another key result of this work.

In contrast, in the S-phase, we see a clear dependence of the spectral function and the ground state energy on $\phi$.
In particular, the functional dependence of the ground state energy follows the short-junction formula $E(\phi) \propto \cos(\phi)$, see \cref{fig:fig4}(a).
The total energy has a minimum at $\phi = 0$, consistent with the $0$-junction behavior of the singlet ground state in SAIM \cite{meden2019_anderson-jospehson_quantum_dot_theory_JourPhysCondMat}. The most noteworthy feature seen in \cref{fig:fig4}(c) is the closing of the proximitized gap in the $\phi \to \pi$ limit, $\Delta^* \to 0$. The layer metalizes and a quasiparticle peak emerges precisely at $\phi=\pi$, with $\mathcal{A}_{11}(0)$ pinned to the noninteracting value.
This behaviour is also reflected in the low energy scaling of the self-energy $-\Sigma_{11}^{\prime\prime} \propto \omega^2$ in the S phase at $\phi=\pi$ in \cref{fig:fig4}(d), which is characteristic of a Fermi liquid. At $\phi=\pi$ the anomalous part of $\mat{\Sigma}_\mathrm{BCS}$ vanishes as $\cos(\phi/2)$, leaving a self-energy of effectively normal-state character. Despite the gap in $\mat{\Sigma}_\mathrm{BCS}$ and even for $U>U_{c2}$ of the isolated Hubbard layer, the DMFT self-consistency still yields a correlated metal solution.

\begin{figure}[t!]
  \includegraphics{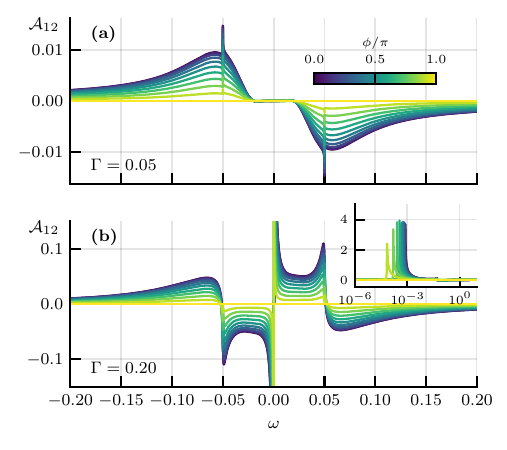}
  \caption{
    Anomalous spectral function $\mathcal{A}_{12}$ for different values of $\phi$ at $U = 3.2$ for
    (a) $\Gamma = 0.05 < \Gamma_{\mathrm{c}}$, and (b) $\Gamma = 0.20 > \Gamma_{\mathrm{c}}$, where $\Gamma_{\mathrm{c}}$ is the transition point between the S- and M-phase at $\phi=0$. Inset in (b) shows data for $\Gamma=0.20$ on a lin-log scale: note that the maximum in $\mathcal{A}_{12}$ is three orders of magnitude bigger in comparison to (a).
  }
  \label{fig:off_diagonal_spectral_function_PHI_DEP_U_3.2_gap_0.05}
\end{figure}

\begin{figure}[h!]
  \includegraphics{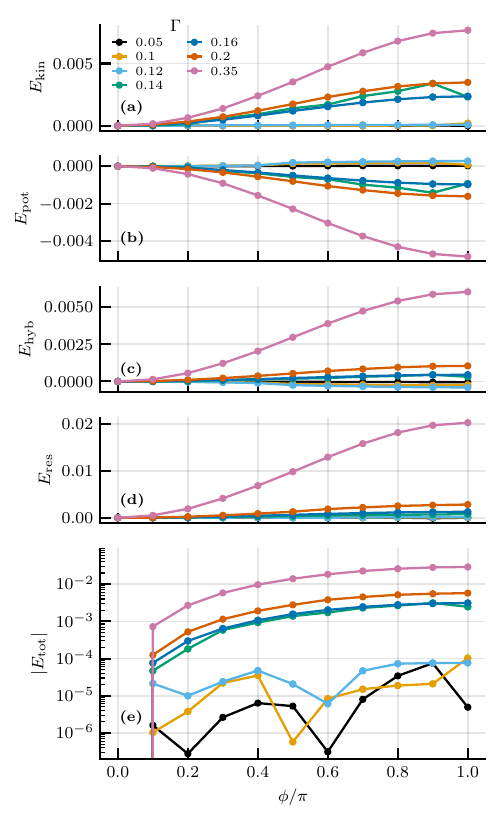}
  \caption{
    Energy of the system as a function of the phase difference $\phi$ at $U = 3.2$ (see Appendix~\ref{sec:energy} for the definitions). From top to bottom: kinetic (\cref{eq:kinetic_energy_lattice}) and potential energies (\cref{eq:interaction_energy_lattice}) of the strongly-correlated layer, hybridization energy (\cref{eq:hybridization_energy_lattice_sc}), SC baths energy (\cref{bath1,bath2}), and the total energy of the system on the log scale, which is the sum of all contributions.
    Note that we plot the energy relative to $E(\phi=0)$.
   For $U=3.2$, $\Gamma_{\mathrm{c}} \approx 0.13$ thus first three lines in the legend correspond to M-phase, the last four to S-phase.
  }
  \label{fig:energies_vs_phi_U_3.2_gap_0.05}
\end{figure}

In \cref{fig:off_diagonal_spectral_function_PHI_DEP_U_3.2_gap_0.05}, we analyse the anomalous spectral function $\mathcal{A}_{12}$ for $U = 3.2$ for the full range of phase bias $\phi$ for (a) $\Gamma=0.05$ (S-phase) and (b) $\Gamma = 0.20$ (M-phase). As we sweep $\phi$ from $0$ to $\pi$, the magnitude of $\mathcal{A}_{12}$ decreases and vanishes at $\phi = \pi$. This can be understood from the form of the off-diagonal component of the superconducting hybridization function, which is proportional to $\cos{\phi/2}$. Note that in panel (b), in the S-phase, the maximum value $\max \mathcal{A}_{12} \approx 4$ (more clearly seen on the full vertical scale in the inset) is more than two orders of magnitude larger than the peak values in the M-phase in panel (a), which are of order 0.01.

In \cref{fig:energies_vs_phi_U_3.2_gap_0.05}, we plot the dependence of energy contributions on the phase bias. The contributions of individual parts of the system to the total energy are derived in Appendix~\ref{sec:energy}. The minimum of the total energy is at $\phi=0$. For $\Gamma < \Gamma_{\mathrm{c}}$, the energy does not depend on $\phi$, as noted in the main text. Due to technical limitations of the DMFT+NRG method for gapped problems (see Appendix~\ref{sec:dep_nrg_params}), this quenching of the phase stiffness in the M-phase can only be established within some numerical uncertainty. More specifically, in the last panel of \cref{fig:energies_vs_phi_U_3.2_gap_0.05} we plot the $\phi$-dependence of the total energy on the logarithmic scale. For $\Gamma>\Gamma_c$, the dispersion is clearly of $E=-E_J \cos(\phi)$ type. For $\Gamma<\Gamma_c$, the results are noisy, and the value of $E_J$ cannot be clearly evaluated. The most {\emph{conservative}} estimation of the upper bound for $E_J$ consists in taking the maximum relative value of total energy. This allows us to show that the suppression of $E_J$ is by at least an order of magnitude; we believe, however, that the suppression is much stronger, but we cannot establish that conclusively using currently available techniques. Since the Josephson current is given directly by the derivative of the energy-phase relation, $I(\phi)=(2e/\hbar) \mathrm{d}E/\mathrm{d}\phi$, these plots directly establish that $I_c$ in the Mott phase is suppressed in the M-phase by the same ratio as the energy-phase relation (i.e., by at least an order of magnitude).

\section{Discussion}\label{sec:discussion}
The two distinct phases in the proximitized Hubbard model can be interpreted from two different perspectives: 1) as extensions of the metallic and Mott insulating phases of the isolated Hubbard region to finite proximity coupling $\Gamma$, and 2) as the singlet and doublet solutions of the effective quantum impurity problem (SAIM) in the DMFT picture. While this connection may appear natural, it is nontrivial: the self-consistency condition significantly modifies the effective hybridization function in the SAIM compared to the conventional impurity problem, with drastic effects. In the quantum dot SAIM, the doublet ground state corresponds to a $\pi$-junction that still carries a sign-reversed Josephson current with somewhat weaker $\phi$-dispersion compared to the singlet ground state\cite{RozhkovArovas1999_PRL_2788, dam2006}. In the lattice problem, the self-consistency instead produces almost complete suppression of the critical current in the M-phase. The Mott charge gap provides a qualitatively different barrier than a band gap in conventional SIS junctions. In the Mott case, the gap is dynamically generated by the self-energy pole structure, and the absence of low-energy quasiparticle states means that the superconducting phase information cannot propagate across the barrier. This distinction between a single-particle gap and a correlation-driven gap is fundamental and should persist beyond single-site DMFT, since it relies on the local self-energy structure. Another striking effect occurs in the $\phi\to\pi$ limit. In the impurity problem, the doublet phase persists to arbitrarily large $\Gamma$ at $\phi=\pi$ (the "doublet chimney" \cite{Bargerbos2022_PRXQuantum_030311,Pavesic2024,Zalom2026}), i.e., the spin cannot be screened. In the correlated junction problem, the proximitized gap in the S-phase collapses and one finds a correlated metal, while the M-phase is unaffected, which is a remnant of the doublet-chimney physics. 

\section{Conclusion}\label{sec:conclusion}

We have analysed the superconducting proximity effect in the Hubbard model in the paramagnetic regime within the DMFT approximation. The calculation of the ground state energy in the junction problem reveals suppression of the critical current in the M-phase.

Several aspects of the model deserve further investigation. We have restricted the analysis to half-filling, where particle-hole symmetry pins the chemical potential and maximizes the Mott gap. Away from half-filling, the Mott gap closes and the system becomes a doped correlated metal; the interplay between doping-induced carriers and the proximity effect is an open question. Nonlocal correlations beyond single-site DMFT on finite-dimensional lattices could modify the quantitative phase boundaries but are not expected to alter the qualitative distinction between M and S phases, which is rooted in the local self-energy structure. Magnetic ordering could introduce additional competing phases.

Our results are directly relevant to recent experimental platforms in van der Waals heterostructures involving Mott insulating layers sandwiched between superconducting contacts. The suppression of the Josephson current in the M-phase could be observed as an unusually small critical current compared to junctions with non-correlated barriers. The coexistence region in the phase diagram implies hysteretic switching between high and low critical current states as a function of parameters that tune $U/\Gamma$, such as pressure or interlayer spacing.

\section*{Acknowledgements}
We acknowledge the support of the Slovenian Research and Innovation Agency (ARIS) under P1-0416, the HPC RIVR consortium and EuroHPC JU for funding this research by providing computing resources of the HPC system Vega at the Institute of Information Science in Maribor, Slovenia.

\section*{Data availability}
Data that support the findings in this paper are available as a Zenodo repository \cite{rolih2026_zenodo_dataset}.

% \appendix*
\appendix

\section{Inhomogeneous DMFT}\label{sec:iDMFT}

Here we show that the inhomogeneous DMFT approach for layered heterostructures with a single active layer (i.e., one layer which is solved self-consistently and connected to non-interacting leads which act as reservoirs) is mathematically equivalent to the model introduced in the main text.
For simplicity, we work with paramagnetic Hubbard model but the results extend trivially to the case with superconducting Nambu matrix structure describe in the following section.

Following Refs.~\cite{helmes2008_kondo_proximity_effect_metal_into_mott_insulator_PhysRevLett,zenia2009_appearance_fragile_fermi_liquids_finite_width_mott_insulators_idmft_PhysRevLett}, a layered heterostructure can be modelled by the following Hamiltonian
\begin{equation}
  \label{eq:layered_hamiltonian}
    \begin{split}
      H = &-\sum_{ij\sigma,\alpha} t_{ij} c_{i\sigma,\alpha}^\dagger c_{j\sigma, \alpha} - \mu \sum_{i\sigma, \alpha} c^\dagger_{i\sigma, \alpha} c_{i\sigma, \alpha} \\ 
      &- V \sum_{i\sigma,\alpha}(c^\dagger_{i\sigma, \alpha}c_{i\sigma,\alpha+1} + \mathrm{h.c.})\\
      &+ \sum_{i, \alpha} U_{\alpha}\left(n_{i\uparrow, \alpha} - \frac{1}{2}\right)\left(n_{i\downarrow,\alpha} - \frac{1}{2}\right).
    \end{split}
\end{equation}
Here, $\alpha$ is an integer plane index $\alpha\in\mathbb{Z}$, and the label $i$ indexes the sites of the two-dimensional square-lattice in each plane.
The operator $c^\dagger_{i\sigma, \alpha}$ creates an electron of spin $\sigma$ at site $i$ on the plane $\alpha$.
In-plane hopping with parameter $t$ is restricted to nearest neighbors only. Similarly, the inter plane hopping parametrized by $V$ is restricted to neighboring planes. Setting $V = t$ would then correspond to a simple cubic lattice.
The interaction parameter $U_{\alpha}$ may vary from plane to plane.

This many-body Hamiltonian can be solved using the self-consistent inhomogeneous DMFT framework as derived by Potthoff and Nolting~\cite{potthoff1999a_surface_metal_transition_Hubbard_PhysRevB,freericks2004_DMFT_inhomogeneous_multilayered_PhysRevB,freericks2006_transport_nanostructures_DMFT_book}. In the DMFT approximation, the self-energy is assumed to be local
\begin{equation}
  \label{eq:iDMFT_self_energy_approximation}
  \Sigma_{ij,\alpha}(\omega) = \delta_{ij} \Sigma_{\alpha}(\omega).
\end{equation}
We have used translational invariance within each plane to write the self-energy as independent of the site index $i$, but it depends on the plane index $\alpha$ since the system is inhomogeneous in that direction.
We also assume a paramagnetic solution and thus drop the spin index.

The non-interacting lattice Green's function for the layered system in \cref{eq:layered_hamiltonian} can be written in the mixed basis of two-dimensional in-plane momentum $\vec{k}^{\parallel} = (k_x, k_y)$ and plane index $\alpha$ as
\begin{widetext}
\begin{equation}
  \label{eq:non_interacting_gf_layered}
  \sum_{\gamma}\left[ (z + \mu - \epsilon^{\parallel}_{\alpha\vec{k}_{\parallel}}) \delta_{\alpha\gamma} + V\delta_{\alpha+1,\gamma} + V\delta_{\alpha-1,\gamma} \right] G_{\gamma\beta}^0(\vec{k}_{\parallel}, z) = \delta_{\alpha\beta}.
\end{equation}
\end{widetext}
Here, $\epsilon^{\parallel}_{\alpha\vec{k}_{\parallel}}$ is the in-plane dispersion relation for a two-dimensional plane $\epsilon^{\parallel}_{\alpha\vec{k}_{\parallel}} = -2t(\cos k_x + \cos k_y)$ from now on taken to be independent of the plane index $\alpha$.
From \cref{eq:non_interacting_gf_layered}, one can see that the inverse of the non-interacting lattice Green's function as a matrix in plane indices is a tridiagonal matrix.

The Dyson equation for the interacting lattice Green's function in real-space is then
\begin{equation}
  \label{eq:dyson_equation_layered}
  \begin{split}
  G_{\alpha\beta}(\vec{k}^\parallel, z) &= G_{\alpha\beta}^0(\vec{k}^\parallel, z) \\
  &+ \sum_{\gamma} G_{\alpha\gamma}^0(\vec{k}^\parallel, z) \Sigma_{\gamma}(z) G_{\gamma\beta}(\vec{k}^\parallel, z)
  \end{split}
\end{equation}
where we have used the locality of the self-energy in \cref{eq:iDMFT_self_energy_approximation}.
The equation for its matrix elements is then
\begin{equation}
  \label{eq:interacting_gf_layered}
  \begin{split}
  &\left[z + \mu - \Sigma_{\alpha} - \epsilon_{\vec{k}^{\parallel}}\right] G_{\alpha\beta}(\vec{k}^{\parallel}, z)\\
  &+ V G_{\alpha+1,\beta}(\vec{k}^{\parallel}, z) + V G_{\alpha-1,\beta}(\vec{k}^{\parallel}, z) \\
  &= \delta_{\alpha\beta}.
  \end{split}
\end{equation}
This is a tridiagonal system of coupled equations for the lattice Green's function $\mat{G}(\vec{k}^{\parallel}, z)$ with indices in $z$-direction $\alpha, \beta \in \mathbb{Z}$ which can be solved directly via continued fraction expansion~\cite{economou2005_greens_functions_quantum_physics_book,freericks2004_DMFT_inhomogeneous_multilayered_PhysRevB}.
Of course, in the actual implementation, one needs to truncate the system to a finite number of planes labeled by $\alpha = 1, 2, \ldots, N$ and impose appropriate boundary conditions at the edges. 
The local Green's function on plane $\alpha$ is expressed in terms of two continued fractions $L_{\alpha}(\vec{k}^{\parallel}, z)$ and $R_{\alpha}(\vec{k}^{\parallel}, z)$ as
\begin{equation}
  \label{eq:local_gf_layered_continued_fraction}
  G_{\alpha\alpha}(\vec{k}^{\parallel}, z) = \frac{1}{L_{\alpha}(\vec{k}^{\parallel}, z) + R_{\alpha}(\vec{k}^{\parallel}, z) - Z_{\alpha} + \epsilon_{\vec{k}^{\parallel}}},
\end{equation}
with $Z_{\alpha}(z) \equiv z + \mu - \Sigma_{\alpha}(z)$.
The left and right continued fractions for $1 \leq \alpha \leq N$ are defined by the following expressions
\begin{widetext}
\begin{align*}
  &L_{\alpha}(\epsilon_{\vec{k}^{\parallel}}, z) = Z_{\alpha}(z) - \epsilon_{\vec{k}^{\parallel}} - 
  \cfrac{V^2}{Z_{\alpha-1}(z) - \epsilon_{\vec{k}^{\parallel}} - 
  \cfrac{V^2}{Z_{\alpha-2}(z) - \epsilon_{\vec{k}^{\parallel}} - 
  \cfrac{\ddots}{Z_1(z) - \epsilon_{\vec{k}^{\parallel}} - 
  \cfrac{V^2}{\frac{1}{2}(z + \mu - \epsilon_{\vec{k}^{\parallel}}) \pm \frac{1}{2}\sqrt{(z + \mu - \epsilon_{\vec{k}^{\parallel}})^2 - 4}}}}},\\
  &R_{\alpha}(\epsilon_{\vec{k}^{\parallel}}, z) = Z_{\alpha}(z) - \epsilon_{\vec{k}^{\parallel}} - 
  \cfrac{V^2}{Z_{\alpha+1}(z) - \epsilon_{\vec{k}^{\parallel}} - 
  \cfrac{V^2}{Z_{\alpha+2}(z) - \epsilon_{\vec{k}^{\parallel}} - 
  \cfrac{\ddots}{Z_N(z) - \epsilon_{\vec{k}^{\parallel}} - 
  \cfrac{V^2}{\frac{1}{2}(z  + \mu - \epsilon_{\vec{k}^{\parallel}}) \pm \frac{1}{2}\sqrt{(z + \mu - \epsilon_{\vec{k}^{\parallel}})^2 - 4}}}}},
\end{align*}
\end{widetext}
where the boundary conditions are set by assuming semi-infinite metallic leads at both ends of the interacting region
\begin{equation}
  L_{0}(\epsilon_{\vec{k}^{\parallel}}, z) = R_{N+1}(\epsilon_{\vec{k}^{\parallel}}, z) = g_{1d}^{-1}(z, \epsilon_{\vec{k}^{\parallel}})
\end{equation}
with
\begin{equation}
  \begin{split}
    g_{1d}^{-1}(z, \epsilon_{\vec{k}^{\parallel}}) &= \frac{1}{2}(z + \mu - \epsilon_{\vec{k}^{\parallel}})\\
    &\pm \frac{1}{2}\sqrt{(z + \mu - \epsilon_{\vec{k}^{\parallel}})^2 - 4V^2};
  \end{split}
\end{equation}
the choice of sign in front of the square root is enforced by causality.

Now, let us consider the case of a single active layer. Using the same boundary conditions of semi-infinite metallic leads on both sides, the matrix inverse of the lattice Green's function from \cref{eq:interacting_gf_layered} is
\begin{equation*}
  \begin{split}
  &\mat{G}^{-1}(\vec{k}^{\parallel}, z) =\\
  &\begin{pmatrix}
    g_{1d}(z-\varepsilon_{\vec{k}^{\parallel}}) & V & 0 \\
    V & z + \mu - \epsilon_{\vec{k}^{\parallel}} - \Sigma_{1}(z) & V \\
    0 & V & g_{1d}(z-\varepsilon_{\vec{k}^{\parallel}})
  \end{pmatrix}
  \end{split};
\end{equation*}
using the Schur complement formula [or directly \cref{eq:local_gf_layered_continued_fraction}] the active layer's local Green's function (local in real-space) is
\begin{equation*}
  G_{11}(\vec{k}^{\parallel}, z) = \frac{1}{z + \mu - \epsilon_{\vec{k}^{\parallel}} - \Sigma_{1}(z) - 2V^2 g_{1d}(z-\varepsilon_{\vec{k}^{\parallel}})}.
\end{equation*}
This is exactly the same as the Green's function of a single-layer Hubbard model with an additional self-energy term originating from the coupling to two semi-infinite metallic leads acting as reservoirs.
The last step is the sum over the in-plane momenta $\vec{k}^{\parallel}$ to obtain the local Green's function on the active layer which is performed as an integral over the non-interacting density of states for the two-dimensional square lattice
\begin{equation}
  \label{eq:local_gf_single_active_layer_integration}
  G_{11}(z) = \int  \frac{\dif \epsilon \rho_{2d}(\epsilon)}{z + \mu - \epsilon - \Sigma_{1}(z) - 2V^2 g_{1d}(z-\epsilon)}.
\end{equation}

Instead of formulating a fully self-consistent inhomogeneous DMFT, we can simply model the effect of the electronic reservoirs via an additional self-energy term in the lattice Green's function (see Ref.~\cite{bag2024_coupling_strongly_correlated_electron_system_elecetronic_reservoir_PhysRevB}).
On the level of the Hamiltonian, this corresponds to adding a coupling to a non-interacting bath to each site of the lattice
\begin{equation}
  H_{i}^{\mathrm{c}} = \gamma \sum_{\vec{q},\sigma} (c_{i\sigma}^\dagger a_{i, \vec{q}\sigma} + \mathrm{h.c.}) + \sum_{\vec{q},\sigma} \varepsilon_{\vec{q}} a_{i, \vec{q}\sigma}^\dagger a_{i, \vec{q}\sigma}.
\end{equation}
Here, $a_{i, \vec{q}\sigma}^\dagger$ creates an electron in the bath coupled to site $i$ with momentum $\vec{q}$ and spin $\sigma$ and $\gamma=V$ is the effective hybridization strength between the lattice site and the bath.

The Green's function of the non-interacting bath is
\begin{equation}
  g_{\mathrm{bath}}(z) = \sum_{\vec{q}} \frac{1}{z - \varepsilon_{\vec{q}}}
\end{equation}
and the effect of the bath on the lattice Green's function is then an additional self-energy term on each lattice site
\begin{equation}
  \Sigma_{\mathrm{bath}}(z) = V^2 g_{\mathrm{bath}}(z).
\end{equation}

In DMFT, the lattice model is mapped onto an effective impurity model. The impurity propagator is
\begin{equation}
  G_{\mathrm{imp}}(z) = \frac{1}{z + \mu - \Delta_{\mathrm{eff}}(z) - \Sigma_U(z)}
\end{equation}
and the self-consistency condition is $G_{\mathrm{imp}}(z) = G_{\mathrm{loc}}(z)$, where $G_{\mathrm{loc}}(z)$ is the local lattice Green's function including the additional self-energy term from the bath.

In paramagnetic DMFT the lattice enters the problem only through its density of states. As concerns the Mott MIT, the detailed form of the DOS matters little \cite{bulla1999_zero_temperature_metal_insulator_transition_nrg_PhysRevLett}. For simplicity, instead of the square lattice DOS we thus take the Bethe lattice DOS. For DMFT calculations on the Bethe lattice, the self-consistency condition simplifies and the effective impurity hybridization function is  
\begin{equation}
  \Delta_{\mathrm{eff}}(z) = t^2 G_{\mathrm{loc}}(z) + \Sigma_{\mathrm{bath}}(z),
\end{equation}
where $t$ is the hopping parameter of the Bethe lattice. In the following, we will use the extension of this approach to the superconducing (SC) case using Nambu formalism.

\section{Green's functions in Nambu space}\label{sec:green_functions}

Defining the Nambu spinors $\psi_{i\lambda, \vec{q}}^{\dagger} = (c_{i\lambda,\vec{q}\uparrow}^\dagger, c_{i\lambda,-\vec{q}\downarrow})$ for the SC region and $\phi_{i}^\dagger = (d_{i\uparrow}^\dagger, d_{i\downarrow})$ for the strongly-correlated region, the Hamiltonian can be written as
\begin{equation}
  H_{\mathrm{sc}, i\lambda}^{\mathrm{c}} = \sum_{\vec{q}} V_{\lambda} (\phi_{i}^\dagger \mat{\tau}_3 \psi_{i\lambda, \vec{q}} + \mathrm{h.c.}) + \sum_{\vec{q}} \psi_{i\lambda, \vec{q}}^\dagger \mat{h}_{\lambda,\vec{q}} \psi_{i\lambda, \vec{q}},
\end{equation}
with $\mat{h}_{\lambda,\vec{q}}$ defined as
\begin{equation}
  \label{eq:h_sc_nambu}
  \mat{h}_{\vec{q}} =
  \begin{pmatrix}
    \varepsilon_{\vec{q}} & \Delta_{\lambda} \\
    \Delta_{\lambda}^{*}            & -\varepsilon_{\vec{q}}
  \end{pmatrix},
\end{equation}
and $\tau_3$ is the Pauli matrix Z.
The Green's function of the superconducting region is given by
\begin{widetext}
\begin{equation}
  \label{eq:gf_bcs_k}
  \mat{g}_{\lambda, \vec{q}}(z) = \left[ z \identity - \mat{h}_{\lambda,\vec{q}} \right]^{-1} = 
  \frac{1}{z^2 - (\varepsilon_{\vec{q}}^2 + |\Delta_{\lambda}|^2)}
  \begin{pmatrix}
    z + \varepsilon_{\vec{q}} & \Delta_{\lambda} \\
    \Delta_{\lambda}^{*}               & z - \varepsilon_{\vec{q}} 
  \end{pmatrix}.
\end{equation}
Assuming a constant DOS $\rho_{\mathrm{SC}}$ in the SC region with half-bandwidth $D_{\mathrm{SC}}$, the integration over momenta $\vec{q}$ gives
\begin{equation}
  \label{eq:gf_bcs}
  \mat{g}_{\lambda}(z) = -\frac{2\rho_{\mathrm{SC}}}{\sqrt{|\Delta_{\lambda}|^2 - z^2}} \arctan\left( \frac{D_{\mathrm{SC}}}{\sqrt{|\Delta_{\lambda}|^2 - z^2}} \right)
  \begin{pmatrix}
    z & \Delta_{\lambda} \\
    \Delta_{\lambda}^{*} & z
  \end{pmatrix}.
\end{equation}
The effect of the SC regions on the Hubbard layer is  an additional self-energy term on each lattice site
\begin{equation}
  \mat{\Sigma}_{\mathrm{BCS}}(z) = \sum_{\lambda} V_{\lambda}^2 \mat{\tau}_3 \mat{g}_{\lambda}(z) \mat{\tau}_3.
\end{equation}
In the case of symmetric coupling, $V_L = V_R = V$, equal superconducting gaps $|\Delta_L| = |\Delta_R| = |\Delta|$ and phases $\phi_L = -\phi_R = \phi/2$, this becomes
\begin{equation}
  \label{eq:sigma_bcs_symmetric}
  \mat{\Sigma}_{\mathrm{BCS}}(z) = -\frac{\Gamma}{\pi}\frac{1}{\sqrt{|\Delta|^2 - z^2}} \arctan\left( \frac{D_{\mathrm{SC}}}{\sqrt{|\Delta|^2 - z^2}} \right)
  \begin{pmatrix}
     z &  -|\Delta| \cos(\phi/2) \\
    - |\Delta| \cos(\phi/2) & z
  \end{pmatrix}.
\end{equation}
\end{widetext}
We have defined the hybridization strength as 
\begin{equation}
\Gamma = 2 \pi \rho_{\mathrm{SC}}(0) V^2.
\end{equation}
In the wide-band limit, $D_{\mathrm{SC}} \to \infty$, and for $\phi=0$, we recover the standard superconducting hybridization function for a single SC contact:
\begin{equation}
  \mat{\Sigma}_{\mathrm{BCS}}(z) = - \frac{\Gamma}{\sqrt{|\Delta|^2 - z^2}}
  \begin{pmatrix}
     z &  -|\Delta| \\
    - |\Delta| & z
  \end{pmatrix}.
\end{equation} 

\section{Effective impurity model, self-consistency}\label{sec:self-consistency}

The interacting lattice Green's function for the Hubbard model in Nambu space is given by
\begin{equation}
  \mat{G}(\vec{k}, \omega) = 
  \begin{pmatrix}
    \llangle d_{\vec{k}\uparrow};  d_{\vec{k}\uparrow}^\dagger \rrangle_{\omega} & \llangle d_{\vec{k}\uparrow};  d_{-\vec{k}\downarrow} \rrangle_{\omega} \\
    \llangle d_{-\vec{k}\downarrow}^\dagger;  d_{\vec{k}\uparrow}^\dagger \rrangle_{\omega} & \llangle d_{-\vec{k}\downarrow}^\dagger;  d_{-\vec{k}\downarrow} \rrangle_{\omega}
  \end{pmatrix}
\end{equation}
with the standard notation $\llangle A; B \rrangle_{\omega} = - \iu \int \e^{\iu\omega t} \Theta(t) \langle [A(t), B] \rangle \dif t$. The off-diagonal (anomalous) elements are non-zero only in the presence of SC symmetry breaking. For the repulsive Hubbard model considered here, the SC symmetry breaking arises from the proximity effect of the SC regions when $\Gamma > 0$. (For attractive Hubbard model with $U<0$, the SC symmetry breaks spontaneously \cite{bauer2009_DMFT_NRG_hubbard_sc_PhysRevB}.)

The interacting Green's function including the effect of the superconducting regions is
\begin{equation}
  \mat{G}^{-1}(\vec{k}, \omega) = \omega \identity - (\varepsilon_{\vec{k}} - \mu) \mat{\tau}_3  - \mat{\Sigma}_{\mathrm{BCS}}(\omega) - \mat{\Sigma}_U(\vec{k}, \omega),
\end{equation}
where $\mat{\Sigma}_{\mathrm{BCS}}$ is given by \cref{eq:sigma_bcs_symmetric} and $\varepsilon_{\vec{k}}$ is the dispersion relation for the Bethe lattice. 

We use dynamical mean-field theory (DMFT) to solve the interacting lattice model.
The self-energy $\mat{\Sigma}_U(\vec{k}, \omega)$ is local 
\begin{equation}
  \mat{\Sigma}_U(\vec{k}, \omega) \approx \mat{\Sigma}_U(\omega).
\end{equation}
This is an exact expression for the Bethe lattice~\cite{georges1996_dmft_RevModPhys}.
The effective impurity model is the superconducting Anderson impurity model (SAIM)
\begin{align}
  H &= H_{\mathrm{imp}} + H_{\mathrm{bath}} + H_{\mathrm{hyb}},\\
  H_{\mathrm{imp}} &= \varepsilon_d \sum_{\sigma} d_{\sigma}^\dagger d_{\sigma} + U n_{\uparrow} n_{\downarrow},\\
  H_{\mathrm{bath}} &= \sum_{\vec{k},\sigma} \varepsilon_{\vec{k}} c_{\vec{k}\sigma}^\dagger c_{\vec{k}\sigma} + \sum_{\vec{k}} \left( \Delta_{\mathrm{bath},\vec{k}} c_{\vec{k}\uparrow}^\dagger c_{-\vec{k}\downarrow}^\dagger + \mathrm{h.c.} \right),\\
  H_{\mathrm{hyb}} &= \sum_{\vec{k},\sigma} V_{\vec{k}} (d_{\sigma}^\dagger c_{\vec{k}\sigma} + \mathrm{h.c.}).
\end{align}
Here, $\epsilon_{\vec{k}}$, $\Delta_{\mathrm{bath},\vec{k}}$ and $V_{\vec{k}}$ are the bath parameters to be determined self-consistently.
The non-interacting Green's function of the impurity model in Nambu space is
\begin{equation}
  \mat{G}_{0,\mathrm{imp}}^{-1}(\omega) = \omega \identity - \varepsilon_d \mat{\tau}_3 - \mat{\Delta}_{\mathrm{eff}}(\omega).
\end{equation}

The effective Weiss field $\mat{\mathcal{G}}_0$ is a matrix in Nambu space.
The DMFT self-consistency equation is
\begin{equation}
  \mat{\mathcal{G}}_0^{-1}(\omega) = \mat{G}_{\mathrm{loc}}^{-1}(\omega) + \mat{\Sigma}(\omega).
\end{equation}
We use the NRG to calculate the self-energy $\mat{\Sigma}_U(\omega)$ of the effective impurity model.
From this we can obtain the local lattice Green's function $\mat{G}_{\mathrm{loc}}(\omega)$ using the Hilbert transform (see next section).
Using the self-consistency condition we can then update the effective hybridization function as
\begin{equation}
  \mat{\Delta}_{\mathrm{eff}}(\omega) = t^2 \mat{\tau}_3 \mat{G}_{\mathrm{loc}}(\omega) \mat{\tau}_3 + \mat{\Sigma}_{\mathrm{BCS}}(\omega).
\end{equation}

The model partly accounts for the back-action of Hubbard layer on the superconductors. It does not include the local renormalization of the superconducting gap, which requires adding ``buffer'' layers with attractive interaction and a self-consistent calculation  \cite{Freericks2001, miller2001, FREERICKS2002,freericks2004_DMFT_inhomogeneous_multilayered_PhysRevB}. Phenomena in buffer layers add further complexity to the problem \cite{Freericks2001} and can be quantitatively important, but should not affect the dynamical correlation effects in the Hubbard layer that are the focus of this work. On the other hand, the inverse proximity effect is fully included in our method, because the calculation is based on an exact mapping with no further approximations, as explained in this section. For this reason, in the section on calculating the total system energy, we will include the corresponding energy terms. In other words, in our approach we take into account the back-action at the level of Green's functions, but not at the level of Hamiltonian parameters.

\section{Hilbert transform in matrix space}\label{sec:hilbert}

Using the Nambu matrix form of the interaction self-energy $\mat{\Sigma}(\vec{k}, \omega)$, the inverse of the interacting lattice Green's function for the proximitized Hubbard model can be written as
\begin{widetext}
\begin{equation}
  \label{eq:lattice_gf_inverse}
  \mat{G}^{-1}(\vec{k}, \omega) = \mat{G}_{0}^{-1}(\vec{k}, \omega) - \mat{\Sigma}(\vec{k}, \omega) = \omega \identity - (\varepsilon_{\vec{k}} - \mu) \mat{\tau}_3  - \mat{\Sigma}_{\mathrm{BCS}} - \mat{\Sigma}_U(\vec{k}, \omega).
\end{equation}
The effect of the SC leads is to add an extra self-energy term $\mat{\Sigma}_{\mathrm{BCS}}$ to the expression of the standard Hubbard model.

The local Green's function is obtained by summing over all momenta $\vec{k}$ for every matrix element
\begin{equation}
  \label{eq:local_gf_hilbert_transform}
  \mat{G}_{\mathrm{loc}}(\omega) = \frac{1}{N} \sum_{\vec{k}} \mat{G}(\vec{k}, \omega) = \int \dif \varepsilon \rho(\varepsilon) \mat{G}(\varepsilon, \omega) = \int \dif \varepsilon \rho(\varepsilon) \left[\omega \identity - (\varepsilon - \mu) \mat{\tau}_3 - \mat{\Sigma}_{\mathrm{BCS}}(\omega) - \mat{\Sigma}_U(\omega)\right]^{-1}.
\end{equation}
Here, $\rho$ is the non-interacting DOS. Following Ref.~\cite{bauer2009_DMFT_NRG_hubbard_sc_PhysRevB}, we first perform partial fraction decomposition of the integrand. We define
\begin{align}
  \zeta_1(\omega) = \omega + \mu - \Sigma_{11}(\omega), \\
  \zeta_2(\omega) = \omega - \mu - \Sigma_{22}(\omega),
\end{align}
where $\Sigma_{ij}$ denotes the component of the matrix $\mat{\Sigma} = \mat{\Sigma}_U + \mat{\Sigma}_{\mathrm{BCS}}$ in row $i$ and column $j$. Entries of the matrix $\mat{G}(\varepsilon, \omega)$ can then be written as
\begin{spreadlines}{1em}
  \begin{align}
    G_{11}(\varepsilon, \omega) =  \frac{\zeta_2 + \varepsilon}{(\zeta_1 - \varepsilon)(\zeta_2 + \varepsilon) - \Sigma_{12}\Sigma_{21}},\\
    G_{12}(\varepsilon, \omega) = \frac{\Sigma_{12}}{(\zeta_1 - \varepsilon)(\zeta_2 + \varepsilon) - \Sigma_{12}\Sigma_{21}},\\
    G_{21}(\varepsilon, \omega) = \frac{\Sigma_{21}}{(\zeta_1 - \varepsilon)(\zeta_2 + \varepsilon) - \Sigma_{12}\Sigma_{21}},\\
    G_{22}(\varepsilon, \omega) = \frac{\zeta_1 - \varepsilon}{(\zeta_1 - \varepsilon)(\zeta_2 + \varepsilon) - \Sigma_{12}\Sigma_{21}},
  \end{align}
\end{spreadlines}
The roots of the denominator are
\begin{equation}
  \label{eq:epsplusmin}
  \varepsilon_\pm = \frac{1}{2}\left(\zeta_1 - \zeta_2 \pm \sqrt{(\zeta_1 + \zeta_2)^2 - 4 \Sigma_{12}\Sigma_{21}}\right)
\end{equation}
and the partial fraction decomposition yields
\begin{spreadlines}{1em}
  \begin{align}
    G_{11}(\varepsilon, \omega) &= \frac{\zeta_2 + \varepsilon_+}{(\varepsilon_+ - \varepsilon_-)(\varepsilon_{+} - \varepsilon)} + \frac{\zeta_2 + \varepsilon_-}{(\varepsilon_- - \varepsilon_+)(\varepsilon_{-} - \varepsilon)} = \frac{A_{11}}{\varepsilon_{+} - \varepsilon} + \frac{B_{11}}{\varepsilon_{-} - \varepsilon}, \\
    G_{12}(\varepsilon, \omega) &= \frac{\Sigma_{12}}{(\varepsilon_+ - \varepsilon_-)(\varepsilon_{+} - \varepsilon)} + \frac{\Sigma_{12}}{(\varepsilon_- - \varepsilon_+)(\varepsilon_{-} - \varepsilon)} = \frac{A_{12}}{\varepsilon_{+} - \varepsilon} + \frac{B_{12}}{\varepsilon_{-} - \varepsilon},\\
    G_{21}(\varepsilon, \omega) &= \frac{\Sigma_{21}}{(\varepsilon_+ - \varepsilon_-)(\varepsilon_{+} - \varepsilon)} + \frac{\Sigma_{21}}{(\varepsilon_- - \varepsilon_+)(\varepsilon_{-} - \varepsilon)} = \frac{A_{21}}{\varepsilon_{+} - \varepsilon} + \frac{B_{21}}{\varepsilon_{-} - \varepsilon},\\
    G_{22}(\varepsilon, \omega) &= \frac{\zeta_1 - \varepsilon_+}{(\varepsilon_+ - \varepsilon_-)(\varepsilon_{+} - \varepsilon)} + \frac{\zeta_1 - \varepsilon_-}{(\varepsilon_- - \varepsilon_+)(\varepsilon_{-} - \varepsilon)} = \frac{A_{22}}{\varepsilon_{+} - \varepsilon} + \frac{B_{22}}{\varepsilon_{-} - \varepsilon}, \\
  \end{align}
\end{spreadlines}
\end{widetext}
where we have defined $A_{ij}$ as
\begin{spreadlines}{1em} 
  \begin{align}
    A_{11} &= \frac{\zeta_2 + \varepsilon_+}{\varepsilon_+ - \varepsilon_-}, \quad A_{12} = \frac{\Sigma_{12}}{\varepsilon_+ - \varepsilon_-},\\
    A_{21} &= \frac{\Sigma_{21}}{\varepsilon_+ - \varepsilon_-}, \quad A_{22} = \frac{\zeta_1 - \varepsilon_+}{\varepsilon_+ - \varepsilon_-}
  \end{align}
\end{spreadlines}
and $B_{ij}$ are obtained from $A_{ij}$ through the replacement $\varepsilon_+ \leftrightarrow \varepsilon_-$
\begin{spreadlines}{1em} 
  \begin{align}
    B_{11} &= \frac{\zeta_2 + \varepsilon_-}{\varepsilon_- - \varepsilon_+}, \quad B_{12} = \frac{\Sigma_{12}}{\varepsilon_- - \varepsilon_+},\\
    B_{21} &= \frac{\Sigma_{21}}{\varepsilon_- - \varepsilon_+}, \quad B_{22} = \frac{\zeta_1 - \varepsilon_-}{\varepsilon_- - \varepsilon_+}.
  \end{align}
\end{spreadlines}

The integral in \cref{eq:local_gf_hilbert_transform} can now be performed using the standard Hilbert transform defined as
\begin{equation}
  \label{eq:hilbert_transform}
  \mathcal{H}[f](z) = \int_{-\infty}^{\infty} \dif \varepsilon \frac{f(\varepsilon)}{z - \varepsilon}.
\end{equation}

The final expressions for the components of the local Green's function are
\begin{align}
  [\mat{G}_\mathrm{loc}(\omega)]_{11} = A_{11} \mathcal{H}[\rho](\varepsilon_+) + B_{11} \mathcal{H}[\rho](\varepsilon_-),\\
  [\mat{G}_\mathrm{loc}(\omega)]_{12} = A_{12} \mathcal{H}[\rho](\varepsilon_+) + B_{12} \mathcal{H}[\rho](\varepsilon_-),\\
  [\mat{G}_\mathrm{loc}(\omega)]_{21} = A_{21} \mathcal{H}[\rho](\varepsilon_+) + B_{21} \mathcal{H}[\rho](\varepsilon_-),\\
  [\mat{G}_\mathrm{loc}(\omega)]_{22} = A_{22} \mathcal{H}[\rho](\varepsilon_+) + B_{22} \mathcal{H}[\rho](\varepsilon_-).
\end{align}
For Bethe lattice DOS, the Hilbert transform can be evaluated analytically as
\begin{equation}
  \label{eq:hilbert_transform_bethe}
  \mathcal{H}[\rho](z) = \frac{1}{2t^2}\left( z - \iu \sign(\mathrm{Im}(z)) \sqrt{4t^2 - z^2} \right).
\end{equation}

\section{Energy of the lattice model}\label{sec:energy}

The expectation value of the Hamiltonian can be expressed in terms of local quantities obtained from the DMFT solution.
It has four contributions: the kinetic energy of the strongly-correlated region, the interaction energy of the strongly-correlated region, the coupling energy between the strongly-correlated region and SC regions, and the energy of the bulk SC regions.

The kinetic energy per site of the strongly-correlated region can be written as~\cite{georges1996_dmft_RevModPhys}
\begin{equation}
  \label{eq:kinetic_energy_lattice}
  \frac{E_{\mathrm{kin}}}{N} = T \Tr \int \dif \varepsilon \rho(\varepsilon) \varepsilon \mat{G}(\varepsilon, \iu\omega_n),
\end{equation}
where $\mat{G}$ is the local Green's function, $\omega_n$ are the fermionic Matsubara frequencies, and the trace is both a Matsubara index summation and a matrix trace in Nambu space.
On the real-frequency axis, this becomes
\begin{equation}
  \label{eq:kinetic_energy_lattice_rf}
  \frac{E_{\mathrm{kin}}}{N} = 2\int_{-\infty}^{\infty} \dif \omega n_f(\omega) \int \dif \varepsilon \rho(\varepsilon) \varepsilon \left(-\frac{1}{\pi}\right) \Im [\mat{G}(\varepsilon, \omega)]_{11},
\end{equation}
where the factor of $2$ accounts for both spin species, $n_f(\omega)$ is the Fermi-Dirac distribution function, and $\mat{G}(\varepsilon, \omega)$ is the full interacting lattice Green's function.

The interacting energy per site of the strongly-correlated region can be calculated directly from the expectation value of the interaction term:
\begin{equation}
  \label{eq:interaction_energy_lattice}
  \frac{E_{\mathrm{int}}}{N} = \frac{U}{N} \sum_{i} \langle n_{i\uparrow} n_{i\downarrow} \rangle = U \langle n_{\uparrow} n_{\downarrow} \rangle.
\end{equation}

For the coupling term between the strongly-correlated region and the SC regions, we need the expression for the mixed Green's function. Using the equation of motion (EOM), this can be expressed as
\begin{equation}
  \mat{G}_{\mathrm{mix}}(\vec{q}, \omega) \equiv \llangle[] \phi_i, \psi^{\dagger}_{i\lambda,\vec{q}} \rrangle_{\omega} = V_{\lambda} \mat{G}_\mathrm{loc} \tau_3 \mat{g}_{\lambda, \vec{q}},
\end{equation}
where $\mat{G}$ is the local Green's function of the strongly-correlated region and $\mat{g}_{\lambda, \vec{q}}$ is the Green's function of the SC region given in \cref{eq:gf_bcs}.
The expectation value is
\begin{equation}
  \frac{E_{\mathrm{hyb}, \mathrm{sc}}}{N} = \sum_{\lambda=\mathrm{L},\mathrm{R}} 2 V_{\lambda}^2 \int_{-\infty}^{\infty} \dif \omega n_f(\omega)\left(-\frac{1}{\pi}\right)\Im\Tr[\mat{G} \tau_3 \mat{g}_{\lambda}\tau_3].
\end{equation}
Using the symmetric choice of $V_{\lambda}$ and $\Delta_{\lambda}$ as before, this becomes
\begin{equation}
  \label{eq:hybridization_energy_lattice_sc}
  \frac{E_{\mathrm{hyb}}}{N} = 2 \int_{-\infty}^{\infty} \dif \omega n_f(\omega) \left(-\frac{1}{\pi}\right) \Im \Tr \left[ \mat{G}_\mathrm{loc}(\omega) \mat{\Sigma}_{\mathrm{BCS}}(\omega) \right].
\end{equation}

The SC region contributions can be divided into normal and anomalous $\langle H_{i\lambda}^{\mathrm{BCS}} \rangle = \langle H_{i\lambda,\mathrm{n}}^{\mathrm{BCS}} \rangle + \langle H_{i\lambda, \mathrm{a}}^{\mathrm{BCS}} \rangle$ as
\begin{align}
  \langle H_{i\lambda,\mathrm{n}}^{\mathrm{BCS}} \rangle &= \sum_{\vec{q},\sigma} \varepsilon_{\vec{q}} \langle c_{i\lambda,\vec{q}\sigma}^\dagger c_{i\lambda,\vec{q}\sigma} \rangle, \\
  \langle H_{i\lambda, \mathrm{a}}^{\mathrm{BCS}} \rangle &= \sum_{\vec{q}} \left(\Delta_{\lambda} \langle c_{i\lambda,\vec{q}\uparrow}^\dagger c_{i\lambda,-\vec{q}\downarrow}^\dagger\rangle + \mathrm{h.c.}\right)
\end{align}
This needs to be evaluated by taking into account the coupling to the strongly-correlated, even if we otherwise neglect the back-action of the interacting layer (the neglect of back-action merely implies that the SC Hamiltonians are fixed, rather than being self-consistently updated, but the inverse proximity effect is fully included in our calculation). The dressed Green's function of the SC reservoir is (using EOMs):
\begin{equation}
    \mat{\tilde{g}}_{\lambda, \vec{q}}(z) = 
    \mat{g}_{\lambda, \vec{q}} 
    + V_{\lambda}^2 \mat{g}_{\lambda, \vec{q}} \mat{\tau}_3 \mat{G}_\mathrm{loc} \mat{\tau}_3 \mat{g}_{\lambda, \vec{q}},
\end{equation}
where $\mat{g}_{\lambda, \vec{q}}$ is the (bare) reservoir Green's function, see \cref{eq:gf_bcs_k}, and $\mat{G}_\mathrm{loc}$ is the local Green's function of the strongly-correlated region. 
Considering only the second term, which arises from the coupling to this region, we get for the normal component
\begin{widetext}
\begin{equation}\label{bath1}
    \langle H_{i\lambda, n}^{\mathrm{BCS}}\rangle = 2V^2_{\lambda} \int \dif\omega n_{\mathrm{F}}(\omega) 
    \int\dif\varepsilon\rho(\varepsilon)\varepsilon
    \left(-\frac{1}{\pi}\right)\mathrm{Im}\,[\mat{g}_{\lambda, \vec{q}} \mat{\tau}_3 \mat{G}_\mathrm{loc} \mat{\tau}_3 \mat{g}_{\lambda, \vec{q}}]_{11}
\end{equation}
and the anomalous component
\begin{equation}\label{bath2}
    \langle H_{i\lambda, a}^{\mathrm{BCS}}\rangle = 2V^2_{\lambda} \int \dif\omega n_{\mathrm{F}}(\omega) 
    \int\dif\varepsilon\rho(\varepsilon)\Delta_{\lambda}
    \left(-\frac{1}{\pi}\right)\mathrm{Im}\,[\mat{g}_{\lambda, \vec{q}} \mat{\tau}_3 \mat{G}_\mathrm{loc} \mat{\tau}_3 \mat{g}_{\lambda, \vec{q}}]_{12}.
\end{equation}

For completeness, we evaluate the constant energy for the isolated SC region (i.e., in the decoupled limit, $V_\lambda \to 0$).
We use Green's functions in \cref{eq:gf_bcs_k} evaluated at fermionic Matsubara frequencies. The first term (normal component) is
\begin{equation}
  \langle H_{i\lambda,\mathrm{n}}^{\mathrm{BCS}} \rangle = \frac{2}{\beta} \sum_{\vec{q}}\sum_{\omega_n} \e^{\iu\omega_n 0^{+}} \varepsilon_{\vec{q}}\frac{\iu\omega_n + \varepsilon_{\vec{q}}}{(\iu\omega_n)^2 - (\varepsilon_{\vec{q}}^2 + |\Delta_{\lambda}|^2)} = -\sum_{\vec{q}} \frac{\varepsilon_{\vec{q}}^2}{\sqrt{\varepsilon_{\vec{q}}^2 + |\Delta_{\lambda}|^2}} \tanh\left(\frac{\beta \sqrt{\varepsilon_{\vec{q}}^2 + |\Delta_{\lambda}|^2}}{2}\right);
\end{equation}
here, $\beta = 1/T$ is the inverse temperature. For flat DOS and $T=0$, this becomes
\begin{align}
  \langle H_{i\lambda,\mathrm{n}}^{\mathrm{BCS}} \rangle &= - \frac{1}{2 D_{\mathrm{SC}}} \int_{-D_{\mathrm{SC}}}^{D_{\mathrm{SC}}} \dif \varepsilon \frac{\varepsilon^2}{\sqrt{\varepsilon^2 + |\Delta_{\lambda}|^2}} = \\
  &= - \frac{1}{2 D_{\mathrm{SC}}} \left[ D_{\mathrm{SC}} \sqrt{D_{\mathrm{SC}}^2 + |\Delta_{\lambda}|^2} - |\Delta_{\lambda}|^2 \arcsinh \left(\frac{D_{\mathrm{SC}}}{|\Delta_{\lambda}|}\right) \right].
\end{align}

The second term (anomalous component) is
\begin{equation}
  \langle H_{i\lambda, \mathrm{a}}^{\mathrm{BCS}} \rangle = \frac{2|\Delta_{\lambda}|^2}{\beta}\sum_{\vec{q}}\sum_{\omega_n}  \frac{\e^{\iu\omega_n 0^+}}{(\iu\omega_n)^2 - (\varepsilon_{\vec{q}}^2 + |\Delta_{\lambda}|^2)}= - \sum_{\vec{q}}\frac{|\Delta_{\lambda}|^2}{\sqrt{\varepsilon_{\vec{q}}^2 + |\Delta_{\lambda}|^2}} \tanh\left(\frac{\beta \sqrt{\varepsilon_{\vec{q}}^2 + |\Delta_{\lambda}|^2}}{2}\right).
\end{equation}
At $T=0$ and for flat DOS, this becomes
\begin{equation}
  \langle H_{i\lambda, \mathrm{a}}^{\mathrm{BCS}} \rangle = - \frac{|\Delta_{\lambda}|^2}{D_{\mathrm{SC}}} \arcsinh\left(\frac{D_{\mathrm{SC}}}{|\Delta_{\lambda}|}\right).
\end{equation}

The final expression for the BCS energy per site (for both regions) is then
\begin{equation}
  \langle H^{\mathrm{BCS}} \rangle = - \sqrt{D_{\mathrm{SC}}^2 + |\Delta|^2} - \frac{|\Delta|^2}{D_{\mathrm{SC}}} \arcsinh\left(\frac{D_{\mathrm{SC}}}{|\Delta|}\right).
\end{equation}
\end{widetext}

%%%%%%%%%%%%%%%%%%%%%%%%%%%%%%%%%%%%%%%%%%%%%%%%%%%%%%%%%%%%%%%%%%%%%%%%%%%%%%%%%%%%%%%%%%%%%%%%%%%%%%%

%\clearpage
%\pagebreak

\section{Surface problem: $\Gamma$-dependence}\label{sec:additional_surface}

%In this appendix, we show supplemental numerical results supporting the findings presented in the main text.

%\subsection{Local spectral function, self-energy and susceptibility}

\begin{figure*}[t!]
  \includegraphics{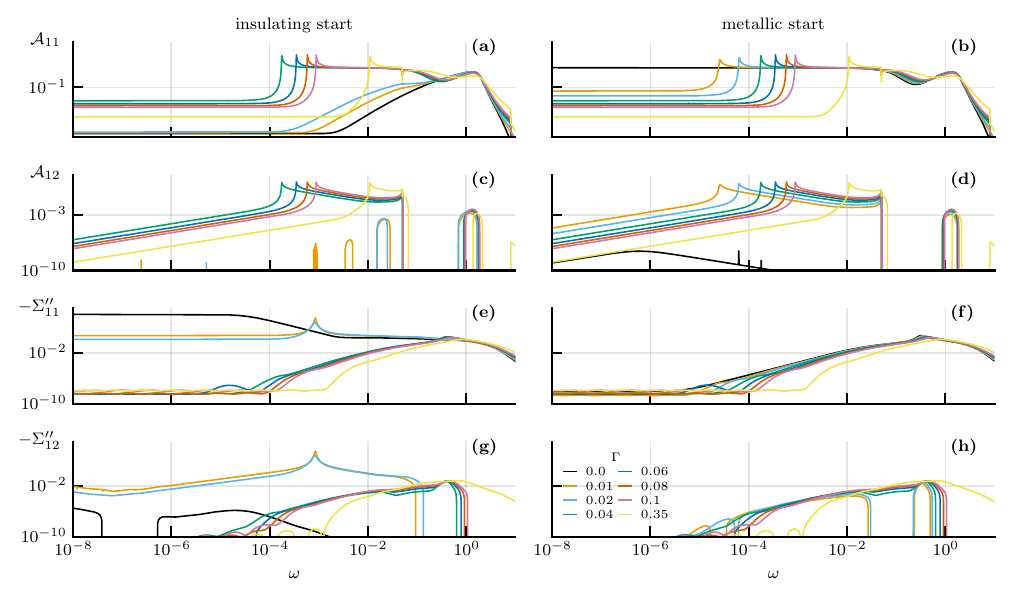}
  \caption{
    Normal (diagonal component in Nambu space) local spectral function, $\mathcal{A}_{11}$, and anomalous (off-diagonal component in Nambu space) local spectral function, $\mathcal{A}_{12}$, as well as the corresponding imaginary parts of the self-energy, $\Sigma''_{11}$ and $\Sigma''_{12}$, for $U=2.6$, $\Delta=0.05$ and $\phi=0$ at various $\Gamma$.
    Left column: results starting from an insulating initial guess. Right column: results starting from a metallic initial guess.
    Note that $U$ belongs to the coexistence region, $U_{c1} < U < U_{c2}$.
  }
  \label{fig:hysteresis_spectral_sigma_U=2.6}
\end{figure*}

In \cref{fig:hysteresis_spectral_sigma_U=2.6}, we show (a,b) the normal spectral function, $\mathcal{A}_{11} \equiv - \mathrm{Im}\, G_{11}/\pi$, and (c, d) the anomalous local spectral function, $\mathcal{A}_{12}  \equiv - \mathrm{Im}\, G_{12}/\pi$, of the strongly-correlated region for $U=2.6>U_{c1}(\Gamma=0)$ at $\phi=0$ for a range of $\Gamma$ for both insulating and metallic initial guesses.
As discussed in the main text, starting from an insulating (more M-like) initial guess, we find a M-phase solution for $\Gamma < \Gamma_c \approx 0.03$ and an S-phase solution for $\Gamma > \Gamma_c$.
Starting from a metallic (more S-like) initial guess, we find the S-phase solution for all values of $\Gamma$.
These findings are consistent with the interpretation that the M-phase is essentially a Mott-insulator solution with weak induced pairing correlations, while the S-phase is essentially a metallic solution with proximitized gap.

\begin{figure*}[t!]
  \includegraphics{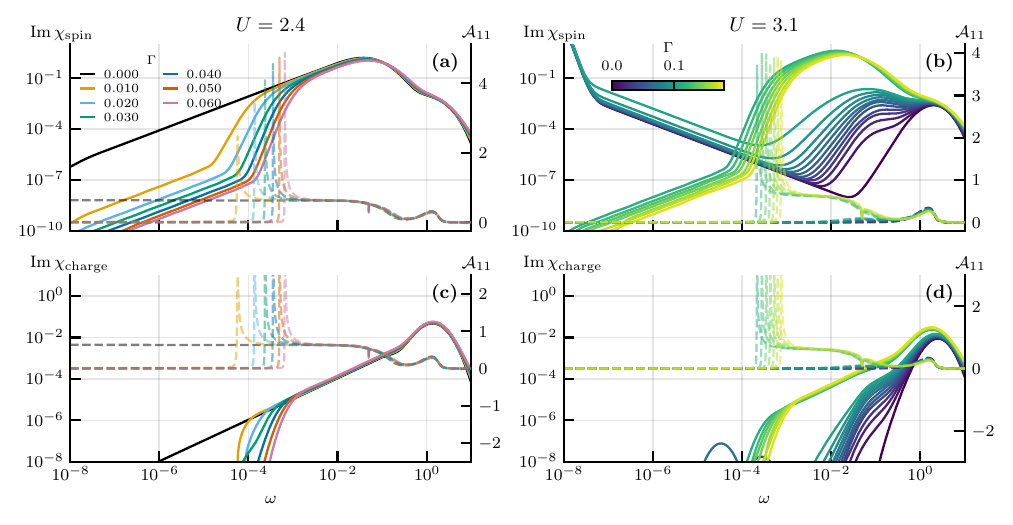}
  \caption{Spin and charge susceptibility for $U = 2.4$ (left column) and $U = 3.1$ (right column) for several values of $\Gamma$.
  For reference we also plot the spectral function $\mathcal{A}_{11}$ (dashed lines).
  }
  \label{fig:susceptibility_charge_spin_U=2.4_U=3.1}
\end{figure*}

In \cref{fig:hysteresis_spectral_sigma_U=2.6}, we also show (e, f) the diagonal and (g, h) the off-diagonal components of the self-energy, $\Sigma^{\prime\prime} = \mathrm{Im}\, \Sigma$, for the same cases.
The self-energy in the insulating state for $\Gamma = 0$ shows a $\delta$-peak at $\omega=0$, characteristic of the Mott insulating state \cite{ZhangRozenbergKotliar1993PRL,RozenbergKotliarZhang1994PRB_MH2,bulla1999_zero_temperature_metal_insulator_transition_nrg_PhysRevLett,Sen2020}.
As $\Gamma$ is increased, the peak is split into two resonances that move away from zero. At the critical value $\Gamma_{\mathrm{c}}$ the peaks disappear completely and the self-energy drops to zero (up to numerical artifacts) for frequencies inside the induced gap.

In \cref{fig:susceptibility_charge_spin_U=2.4_U=3.1}, we show local spin and charge susceptibilities for two different values of the interaction strength: $U=2.4$, which is never in the M phase, and $U=3.1$, which shows a transition at $\Gamma_c \approx 0.11$. The plots complement Fig.~3(a,c) from main text. Note that they are here shown on a log-log scale and without the $1/\omega$ factor.

%Namely, the $E(\phi)$ functions are sine functions, thus the amplitude of energy variation vs. $\phi$ is directly proportional to the critical current $I_c$.

\section{Dependence of the results on technical NRG and DMFT parameters}\label{sec:dep_nrg_params}

Unless stated otherwise, we choose the discretization parameter $\Lambda=4$, average the results over four discretization grids $N_z = 4$ and keep the states with energies up to $E = 12$ in units of rescaled energies, but never less than \num{600} states. This holds for the calculations presented in the main text as well as for the benchmarks discussed in this section.

To obtain the effective bath parameters for the impurity model, we use the discretization scheme described in Ref.~\cite{liu2016_channel_mixing_PhysRevB}.
This approach allows one to discretize general hybridization functions with off-diagonal components in Nambu space. Our implementation of this algorithm in Julia programming language is available from the GitHub repository~\cite{rolih_channelmixingdiscretization}.

Due to its nature, NRG performs best for ungapped hybridization functions, where the discretization yields Wilson chains with conventional asymptotic behavior of hopping constants (monotonic exponentially decreasing as $\Lambda^{-n/2})$. Frequency ranges where hybridization function has low values (i.e., spectral gaps) lead instead to unusual scaling of Wilson chain coefficients that break energy-scale separation, potentially resulting in artifacts in the computed spectra. For this reason, but also to fix possible numerical causality violations, one needs to perform ``clipping'', i.e., restoring the values of self-energy or the hybridization functions to appropriate minimal values of correct sign. The threshold value needs to be chosen carefully. A too low clipping value may result in uncontrolled artifacts, while a too high clipping value will introduce spurious density of states in the gaps. A compromise must be made. Consequently, one can never obtain a truly gapped solution with zero density of states inside the gap, but instead some residual, and potentially noisy, values. In this work, we have implemented the clipping at the level of self-energy.
For a matrix-valued self-energy $\mat{\Sigma}$ we define its spectral representation as
\begin{equation}
    \mat{\mathcal{A}_{\mat{\Sigma}}}(\omega) = -\frac{1}{2\pi\iu}\left[\mat{\Sigma}(\omega + \iu \eta) - \mat{\Sigma}^{\dagger}(\omega + \iu \eta) \right].
\end{equation}
Causality requires that the matrix $\mat{\mathcal{A}}_{\mat{\Sigma}}(\omega)$ is positive semi-definite for every $\omega$. Equivalently, its eigenvalues $\lambda_i(\omega)$ are non-negative. We perform the clipping by diagonalizing $\mat{\mathcal{A}}_{\mat{\Sigma}}$ and requiring $\lambda_i \ge \epsilon_{\mathrm{clip}}$, where $\epsilon_{\mathrm{clip}}$ is the clipping value, typically set to $10^{-7}$. After obtaining the clipped spectral representation, we can reconstruct the value of the self-energy using Kramers-Kronig transformation
\begin{equation}
    \mat{\Sigma}^{\mathrm{clip}}(\omega) = \int \dif \omega^{\prime} \frac{\mat{\mathcal{A}}_{\mat{\Sigma}}^{\mathrm{clip}}(\omega^\prime)}{\omega + \iu\eta - \omega^{\prime}} + \mat{\Sigma}(\infty);
\end{equation}
the last term is the constant Hartree shift of the self-energy.

Another reason for obtaining finite density of states inside the gapped regions is spectral broadening. The raw results in NRG consist of $\delta$ peaks that need to be broadened into a continuous spectral function. Typically one uses a log-Gaussian function with a width that is proportional to frequency. The overall width is controlled by parameter $\alpha$. To be able to reduce $\alpha$, one makes use of twist averaging, i.e., one performs the NRG calculation for several interleaved discretization grids, then the results are averaged. This is controlled by parameter $N_z$, the number of interleaved grids. The method works best when $N_z$ is a power of 2, with commonly used values 2, 4, and 8, which represent a good compromise between the improvement in spectral resolution and computational requirements.

To assess the expected magnitude of artifacts due to clipping and broadening, we perform benchmark calculations with variable NRG parameters. The key finding is that the NRG artifacts are similar as finite-temperature effects. The effective temperature is much lower than the spectral gap, thus in practical calculations the gaps can be unambiguously identified and quantified with no difficulty.

\begin{figure}[t!]
    \includegraphics{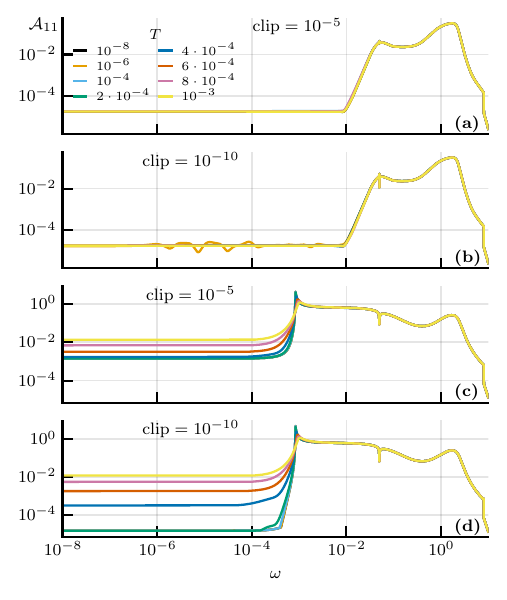}
    \caption{
        Spectral function dependence on temperature and the clipping of the bulk BCS hybridization function. Model parameters: $U=3.2$, hybridization strength (a, b) $\Gamma = 0.05 < \Gamma_{\mathrm{c}}$, (c, d) $\Gamma = 0.2 > \Gamma_{\mathrm{c}}$.
    }
    \label{fig:fig11_temperature_clip_dependence}
\end{figure}

\begin{figure}[b!]
    \includegraphics{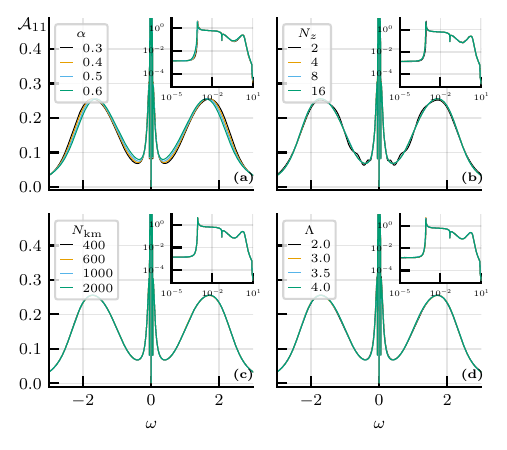}
    \caption{
      Spectral function dependence on the broadening parameter $\alpha$, the number of discretization grids $N_z$, the minimum value of states kept in NRG $N_{\mathrm{km}}$ and discretization parameter $\Lambda$. If not specified in the legend for their respective plots, the NRG parameters are $\alpha=0.6$, $N_{z}=4$, $N_{\mathrm{km}}=600$, $\Lambda=2$. Model parameters: $U=3.2$, $\Gamma = 0.2$.
    }
    \label{fig:fig12_nrg_parameters_dependence}
\end{figure}

In \cref{fig:fig11_temperature_clip_dependence}, we first show the dependence of the local spectral function on (true) temperature and clipping of the superconducting self-energy $\mat{\Sigma}_{\mathrm{BCS}}$ for (a, b) $\Gamma=0.05$ and (c, d) $\Gamma=0.20$. The values at low frequencies do not depend on temperatures as long as $T \ll \Delta$. We find some differences between the two cases. In the $S$-phase the deviations from the $T=0$ results are observed starting from $T\approx4\cdot 10^{-4}$ for clipping parameter $10^{-5}$. By reducing the value of the clipping parameter, $\mathcal{A}_{11}$ values in the gap can be reduced by several orders of magnitude, see panel (d). In the M-phase the saturated values are higher and cannot be much reduced.

We also checked how the result change with the broadening parameter $\alpha$ and the number of discretization grids $N_z$. At low frequencies (see insets of \cref{fig:fig12_nrg_parameters_dependence}) the result are the same for all parameter values; at higher $\alpha$ the superconducting coherence peak becomes less sharp. For small values of $N_z$ we observe oscillatory artifacts in $\mathcal{A}_{11}$ -- this can be fixed by using higher value of $\alpha$ at the cost of overbroadening. All other results in this paper were computed with $\alpha=0.3$ and $N_z=4$.

To update the hybridization function $\mat{\Delta}$ between successive DMFT iterations we used linear mixing to stabilize convergence. This was especially important close to phase transition. For values $(U, \Gamma)$ close to the lines $U_{\mathrm{c}1}$ and $U_{\mathrm{c}2}$ in \cref{fig:fig1_device_phase_diagram_sketch}, we typically mixed 10 \% of the new $\mat{\Delta}$ and needed several hundred iterations for fully converged solutions.

\begin{figure}[t!]
    \centering
    \includegraphics{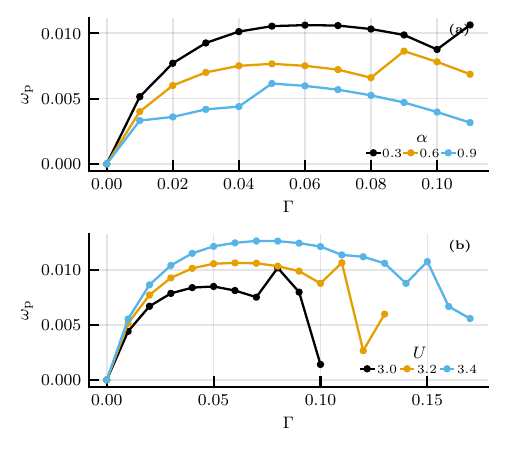}
    \caption{Position of the $\delta$-peak in the self-energy $\Sigma_{11}^{\prime\prime}$ as a function of $\Gamma$. (a) Coulomb repulsion $U = 3.2$. The differences here represent broadening artifacts: a wide log-Gaussian broadening kernel shifts spectral peaks towards lower values. The discontinuous features are also technical artifacts. (b) Results for a range of $U$. Broadening parameter $\alpha = 0.3$. The differences for low-$\Gamma$ represent physical effects, while at higher $\Gamma$ we observe artifacts at $U$-dependent positions.}
    \label{fig:fig13_Sigma_delta_peak_broadening}
\end{figure}

Close to the transition, calculating the position of the $\delta$-peak in the self-energy is numerically sensitive. In \cref{fig:fig13_Sigma_delta_peak_broadening}(a), we show the dependence of $\omega_{\mathrm{p}}$ on the broadening parameter $\alpha$ at $U=3.2$. In \cref{fig:fig13_Sigma_delta_peak_broadening}(b), we plot $\omega_{\mathrm{p}}$ at $\alpha=0.3$ for multiple values of $U$.

\bibliography{references}

\end{document}